\documentclass[aps,twocolumn,amsmath,amssymb]{revtex4-2}
\usepackage[dvipdfmx]{graphicx} 
\usepackage{dcolumn}
\usepackage{bm}
\usepackage{braket}
\usepackage{multirow}
\usepackage[usenames]{color}
\usepackage{hyperref}
\usepackage{bbm}
\hypersetup{
colorlinks = true, 
urlcolor = blue, 
linkcolor = blue, 
citecolor = blue
}

\begin{document}

\title{
Spin-fluctuation-mediated chiral $d\!+\!id'$-wave superconductivity in the $\alpha$--$\mathcal{T}_3$ lattice with an incipient flat band
}

\author{Masataka Kakoi}
\email{kakoi@presto.phys.sci.osaka-u.ac.jp}
\affiliation{Department of Physics, The University of Osaka, Toyonaka, Osaka 560-0043, Japan}

\author{Kazuhiko Kuroki}
\email{kuroki@presto.phys.sci.osaka-u.ac.jp}
\affiliation{Department of Physics, The University of Osaka, Toyonaka, Osaka 560-0043, Japan}

\date{\today}

\begin{abstract}
    We study anisotropic superconductivity in the nearly quarter-filled $\alpha$--$\mathcal{T}_3$ lattice.
    We analyze an extended Hubbard model with off-site attractive interactions within the mean-field framework and find two distinct chiral $d+id'$-wave superconducting phases characterized by different Chern numbers.
    We further investigate the superconducting mechanism mediated by spin fluctuations arising from purely repulsive interactions by applying the fluctuation-exchange (FLEX) approximation to the Hubbard model.
    The gap symmetry obtained by solving the linearized Eliashberg equation is $d$-wave, which corresponds to a $d+id'$-wave superconducting state with a Chern number of $8$, including the spin degree of freedom.
    The $\bm{q}=\bm{0}$ antiferromagnetic spin fluctuation, which possesses the largest spectral weight at finite energies arising from the incipient flat band, gives rise to an effective spin-singlet pairing glue between rim sites.
\end{abstract}

\maketitle

\section{Introduction}

Chiral superconductivity, described by a complex linear combination of multicomponent order parameters, is an exotic superconducting state that spontaneously breaks time-reversal symmetry~\cite{Kallin_2016_review,Ghosh_2021_review}.
It is characterized by a nonzero bulk topological invariant, exhibiting rich properties such as gapless edge states and Majorana quasiparticles~\cite{Sato_2017_review}, and has also attracted attention for its potential applications in quantum computing~\cite{Nayak_2008_review}.

The hexagonal $D_{6h}$ crystal symmetry group has two-dimensional irreducible representations $E_{2g}$ and $E_{1u}$, which are spanned by the even-parity ($d_{x^2-y^2}$, $d_{xy}$) and odd-parity ($p_x$, $p_y$) basis functions, respectively.
Chiral superconductivity often arises when the superconducting order parameter belongs to a higher-dimensional irreducible representation of the crystal point group, making such systems promising candidates for chiral $d$- and $p$-wave superconductivity.
In particular, honeycomb lattice systems, such as graphene, have been theoretically studied for anisotropic superconductivity driven by many-body interactions~\cite{Kuroki_2001,Black-Schaffer_2007,Honerkamp_2008,Gonzalez_2008,Kuroki_2010,Raghu_2010,Pathak_2010,T-Ma_2011,Nandkishore_2012,Nandkishore_2014,Black-Schaffer_2014_review}.
Different approaches have consistently indicated that the $d+id'$-wave superconducting state is favored near the van Hove singularity.

In monolayer graphene, however, superconductivity has not been observed so far, partly because electron correlations are relatively weak.
The situation changes in systems with flat bands, where electron correlations are effectively enhanced. 
Indeed, superconductivity has been realized in twisted bilayer graphene and rhombohedral multilayer graphene~\cite{Y-Cao_2018,Zhou_2021}, and a recent experiment on rhombohedral graphene has reported signatures of chiral superconductivity~\cite{Han_2025}.
Regarding the relationship between flat bands and superconductivity, it has also been investigated that the {\it incipient} flat band coexisting with dispersive bands can promote superconductivity~\cite{Kuroki_2005,Kobayashi_2016,Aida_2024,Aoki_2020_review,*Aoki_2025_review}.

\begin{figure}[t]
\begin{center}
    \includegraphics[width=\columnwidth]{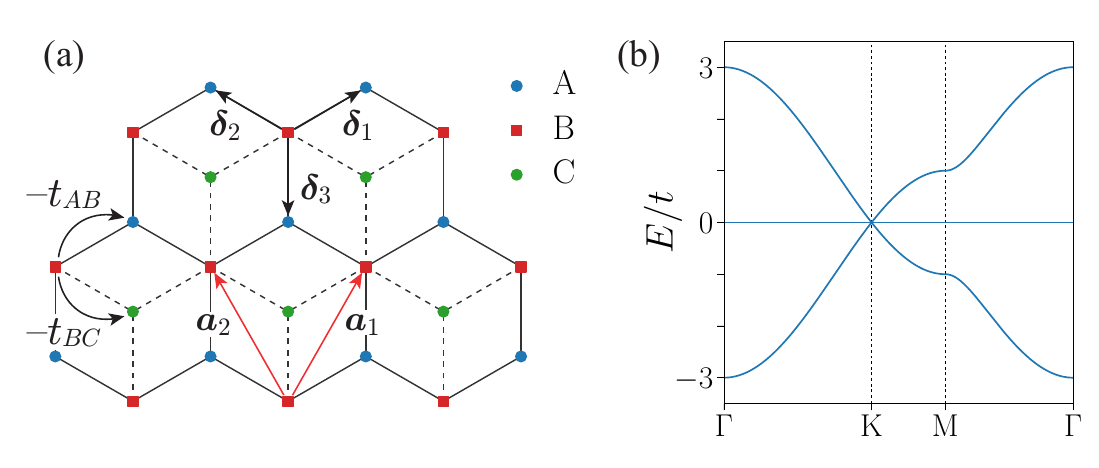}
    \caption{
    {(a)~Real-space structure of the $\alpha$--$\mathcal{T}_3$ lattice. 
    $\bm{\delta}_\ell$ ($\ell=1,2,3$) is given in Eq.~(\ref{eq:AB-vectors}). 
    The lattice vectors are $\bm{a}_1 = \bm{\delta}_1-\bm{\delta}_3$, $\bm{a}_2 = \bm{\delta}_2-\bm{\delta}_3$.
    Hopping is allowed only between the hub (B) and rim (A and C) sites.
    (b)~Tight-binding band structure of the $\alpha$--$\mathcal{T}_3$ lattice.
    The band structure is independent of the hopping ratio $\alpha=t_{\rm AC}/t_{\rm AB}$.}
    } 
    \label{fig:alpha-T3_lattice_and_band}
\end{center}
\end{figure}

Flat-band physics can be explored using lattice models.
Two canonical examples of tight-binding systems that exhibit flat bands are the dice ($\mathcal{T}_3$) lattice~\cite{Horiguchi_1974,Sutherland_1986} and the Lieb lattice~\cite{Lieb_1989}.
These lattices can be decomposed into sublattices with different numbers of sites, and the resulting flat bands are protected by sublattice symmetry. 
In other words, as long as the local topology of the sublattice connectivity is preserved, the flat-band energies in these models do not depend on the precise values of the hopping amplitudes and remain robust in the presence of a magnetic field~\cite{Aoki_1996,Vidal_1998,Leykam_2018_review}.
The $\alpha$--$\mathcal{T}_3$ lattice~\cite{Raoux_2014} is obtained by varying the hopping ratio ($\alpha$) between the hub and rim sites of the dice lattice [see Fig.~\ref{fig:alpha-T3_lattice_and_band}(a)].
This model continuously interpolates between the dice ($\alpha = 1$) and honeycomb ($\alpha = 0$) lattices, and exhibits an $\alpha$-dependent Berry phase.
A variety of nontrivial transport phenomena, such as unconventional magnetic and optical responses, have been theoretically predicted~\cite{Raoux_2014,Piechon_2015,Illes_2015,Illes_2016,Illes_2017,Biswas_2016,Alam_2019,YR-Chen_2019,Day_2019,Iurov_2019,Iurov_2020,JJ-Wang_2020,H-Tan_2020,X-Zhou_2021,J-Wang_2021,F-Li_2022,Tamang_2023,SQ-Lin_2023,HL-Liu_2023,R-Li_2023,Iurov_2023,Abdi_2024,YJ-Wei_2024,Parui_2024,KW-Lee_2024,KW-Lee_2025,Saleem_2025}.
In addition, the presence of a flat band makes electron correlations prominent, making this system an attractive platform for exploring many-body effects.

In this paper, we investigate correlation-driven superconductivity in a nearly quarter-filled $\alpha$--$\mathcal{T}_3$ lattice.
We first analyze an extended Hubbard model that incorporates an on-site repulsion and off-site attraction between neighboring sites. 
We emphasize that the attractive extended Hubbard model on a square lattice has been studied as an effective model for cuprate superconductors, in which the on-site repulsion and nearest-neighbor attraction can account for the $d$-wave superconducting phase emerging near the antiferromagnetic phase~\cite{Micnas_1990_review}. 
We find that the nearest-neighbor attraction in a dice lattice network leads to a $d+id'$-wave superconducting state. 
Interestingly, by changing the strength of the attractive interactions between inequivalent sublattice sites, we identify two distinct topological superconducting phases, each with a different Chern number ($|\mathcal{C}|=4$ and $|\mathcal{C}|=8$ including spin).

We further study the repulsive Hubbard model, which provides a more realistic description of electron systems, using the fluctuation-exchange (FLEX) approximation. 
We show that the incipient flat band enhances finite-energy spin fluctuations at $\bm{q}=\bm{0}$, resulting in a $d+id$-wave superconducting state with a Chern number $|\mathcal{C}| = 8$. 
Interestingly, our results indicate that the model with nearest-neighbor attractive interaction studied in the early half, in the $|\mathcal{C}| = 8$ regime, can be considered an effective model for the on-site repulsive model.

\section{Model}

We study an (extended) Hubbard model of electrons,
\begin{align}\label{eq:TB_Ham}
  \hat{H} = 
  - \!\!\sum_{\braket{i,j},\sigma}\! t_{ij}\, \hat{c}^{\dagger}_{i\sigma} \hat{c}_{j\sigma}
  + \hat{H}_{\rm int},
\end{align}
where $i(j)$, $\sigma$, and $t_{ij}$ denote the site, spin, and hopping amplitude, respectively. 
The two-body interaction term $\hat{H}_{\rm int}$ will be introduced later.
The tight-binding Hamiltonian is defined on the $\alpha$--$\mathcal{T}_3$ lattice~\cite{Raoux_2014}, and its real-space structure is shown in Fig.~\ref{fig:alpha-T3_lattice_and_band}(a).  
The $\alpha$--$\mathcal{T}_3$ lattice is characterized by the  nearest-neighbor hoppings between the hub (B) and rim (A and C) sites:
\begin{equation}\label{eq:hoppings}
t_{ij} = \left\{
\begin{array}{ll}
    t_{\rm AB}=t\cos\varphi, & \text{A--B nearest neighbors},\\[1pt]
    t_{\rm BC}=t\sin\varphi, & \text{B--C nearest neighbors},\\[1pt]
    0, & \text{otherwise}.
\end{array}
\right.
\end{equation}
We define the hopping ratio by $\alpha = t_{\rm BC}/t_{\rm AB} = \tan\varphi$ ($0 \le \alpha \le 1$) for convenience.  
$\alpha = 1$ corresponds to the dice ($\mathcal{T}_3$) lattice, while $\alpha = 0$ corresponds to the honeycomb lattice with isolated C sites.  
The position vectors of the A sites measured from a B site are denoted as
\begin{equation}\label{eq:AB-vectors}
    \bm{\delta}_1 = a\left(\tfrac{\sqrt{3}}{2}, \tfrac{1}{2}\right),\quad 
    \bm{\delta}_2 = a\left(-\tfrac{\sqrt{3}}{2}, \tfrac{1}{2}\right),\quad 
    \bm{\delta}_3 = a(0, -1),
\end{equation}
and those of the C sites are given by $-\bm{\delta}_\ell$.

The tight-binding Hamiltonian is block-diagonal with respect to the Bloch wave vector $\bm{k}$ and can be written as a $3\times3$ matrix corresponding to the three sublattice sites,
\begin{equation}\label{eq:alpha-T3_Ham_k-space}
    H_0(\bm{k}) = 
    \begin{pmatrix}
        0 & f(\bm{k}) \cos\varphi & 0 \\
        f^*(\bm{k}) \cos\varphi & 0 & f(\bm{k}) \sin\varphi \\
        0 & f^*(\bm{k}) \sin\varphi & 0
    \end{pmatrix}.
\end{equation}
Here,
\begin{equation}\label{eq:def_fk}
    f(\bm{k}) = -t\sum_{\ell=1}^3 e^{i\bm{k}\cdot\bm{\delta}_\ell}
\end{equation}
is the structure factor.
The eigenvectors of the Hamiltonian~(\ref{eq:alpha-T3_Ham_k-space}) are given by
\begin{equation}\label{eq:orb2band}
    \ket{\bm{k},\pm} = \frac{1}{\sqrt{2}}
    \begin{pmatrix}
        e^{i\theta(\bm{k})} \cos\varphi\\
        \pm1\\
        e^{-i\theta(\bm{k})} \sin\varphi
    \end{pmatrix}\!,\ \ 
    \ket{\bm{k},0} =
    \begin{pmatrix}
        e^{i\theta(\bm{k})} \sin\varphi\\
        0\\
        -e^{-i\theta(\bm{k})} \cos\varphi
    \end{pmatrix}\!,
\end{equation}
where we define $\theta(\bm{k}) = \mathrm{arg}[f(\bm{k})]$.
The corresponding eigenenergies are
\begin{equation}\label{eq:bands}
    E_{\pm}(\bm{k}) = \pm |f(\bm{k})|,\quad E_0(\bm{k}) = 0.
\end{equation}
The band structure is shown in Fig.~\ref{fig:alpha-T3_lattice_and_band}(b) and is independent of $\alpha$.
For a closed path $C$ enclosing the $\textrm{K}$($\textrm{K}'$) point, the dispersive bands have a Berry phase of $\gamma[C] = \pm\pi\cos2\varphi$, while the flat band has $\gamma[C] = \mp2\pi\cos2\varphi$~\cite{Raoux_2014}.

The model Hamiltonian can potentially be realized in several physical systems.
The dice lattice network emerges in trilayer structures of cubic lattices grown along the $[111]$ crystallographic direction, such as transition-metal oxide (TMO) trilayer heterostructures~\cite{F-Wang_2011,Okamoto_2025_review}.
In addition, Hg$_{1-x}$Cd$_x$Te (MCT) at the critical doping can be mapped onto the $\alpha$--$\mathcal{T}_3$ lattice with $\alpha=1/\sqrt{3}$~\cite{Orlita_2014,Malcolm_2015}.
A recent ARPES experiment has observed the flat bands of the dice lattice limit in the van der Waals electride YCl~\cite{Geng_2026}.
Moreover, theoretical studies suggest that the $\alpha$--$\mathcal{T}_3$ model can be implemented with ultracold atoms in optical lattices~\cite{Rizzi_2006,Bercioux_2009,Raoux_2014}.

\section{Results of the many-body calculations}

\subsection{Mean-field analysis of the extended Hubbard model with off-site attractive interactions \label{sec:mean-field}}

To investigate the anisotropic superconductivity in the $\alpha$--$\mathcal{T}_3$ lattice within the mean-field framework, we study an extended Hubbard model,
\begin{equation}\label{eq:extended_Hubbard_int}
    \hat{H}_{\rm int} = U\sum_{i} \,\hat{n}_{i,\uparrow} \hat{n}_{i,\downarrow}+\frac12\!\sum_{\braket{i,j}, \sigma,\sigma'}\!\! V_{ij} \,\hat{n}_{i,\sigma} \hat{n}_{j,\sigma'},
\end{equation}
where $U(>0)$ is the on-site Coulomb interaction, and 
\begin{equation}
V_{ij} = \left\{
\begin{array}{ll}
    -V_{\rm AB}, & \text{A--B nearest neighbors},\\[1pt]
    -V_{\rm BC}, & \text{B--C nearest neighbors},\\[1pt]
    -V_{\rm CA}, & \text{C--A nearest neighbors},
\end{array}
\right.
\end{equation}
are interactions between nearest-neighbor sites, which we assume to be attractive ($V_{ij}<0$).
We adopt this model because our objective is to investigate anisotropic superconductivity driven by electron correlations. Spin fluctuations can give rise to an effective momentum-dependent pairing interaction. In phenomenological lattice models, this effect is often mimicked by introducing off-site attractive interactions. Indeed, the combination of on-site repulsion and off-site attraction has been widely employed to describe the superconducting ground state of cuprate superconductors~\cite{Micnas_1990_review}.
When the interaction between nearest-neighbor sites is repulsive, superconductivity does not emerge on the square lattice at the mean-field level~\cite{Micnas_1990_review}. (Note that this does not rule out superconductivity arising from generic repulsive interactions.)
On-site attraction ($U<0$) typically stabilizes $s$-wave superconductivity~\cite{Micnas_1990_review}.

Since we discuss spin-singlet superconductivity, considering only the scattering between electrons in the single-particle states $\ket{\bm{k},\uparrow}$ and $\ket{-\bm{k},\downarrow}$ in Eq.~(\ref{eq:extended_Hubbard_int}) leads to
\begin{equation}
    \hat{H}_{\rm int} =\!\!\sum_{\bm{k},\bm{k}',\alpha,\beta}\!\! V_{\alpha\beta}(\bm{k}\!-\!\bm{k}')\,
    \hat{c}^{\dag}_{\bm{k},\uparrow,\alpha} 
    \hat{c}^{\dag}_{-\bm{k},\downarrow,\beta}
    \hat{c}^{}_{-\bm{k}',\downarrow,\beta} 
    \hat{c}^{}_{\bm{k}',\uparrow,\alpha}
\end{equation}
where $\alpha$($\beta$) denotes the sublattice and 
\begin{equation}
V_{\alpha\beta}(\bm{k}) = \left\{
\begin{array}{ll}
    U, & \alpha=\beta,\\[1pt]
    -V_{\alpha\beta}f(\bm{k}), & \alpha\beta\in\{{\rm AB,\,BC,\,CA}\},\\[1pt]
    -V_{\alpha\beta}f^*(\bm{k}), & \alpha\beta\in\{{\rm BA,\,CB,\,AC}\}.
\end{array}
\right.
\end{equation}
In a superconducting state, the expectation value $\braket{\hat{c}^{}_{\bm{k},\uparrow,\alpha}\,\hat{c}^{}_{-\bm{k},\downarrow,\beta} }$ is nonzero.
We therefore introduce the pairing potential $\Delta_{\alpha\beta}(\bm{k})$ defined as
\begin{equation}\label{eq:def_gap_func}
    \Delta_{\alpha\beta}(\bm{k}) = \sum_{\bm{k}'} V_{\alpha\beta}(\bm{k}\!-\!\bm{k}')
    \braket{
    \hat{c}^{}_{-\bm{k}'\!,\downarrow,\beta} \,\hat{c}^{}_{\bm{k}'\!,\uparrow,\alpha}
    }
\end{equation}
and applying the mean-field approximation to decouple the quartic interaction term yields
\begin{equation}
    \hat{H}_{\rm MF} = \sum_{\bm{k}} \hat{\Psi}^{\dag}_{\!\bm{k}} H_{\rm BdG}(\bm{k}) \hat{\Psi}^{}_{\!\bm{k}},
\end{equation}
where 
\begin{equation}\nonumber
    \hat{\Psi}_{\!\bm{k}}=\big(
    \hat{c}^{}_{\bm{k},\uparrow,\rm A},\hat{c}^{}_{\bm{k},\uparrow,\rm B},\hat{c}^{}_{\bm{k},\uparrow,\rm C},
    \hat{c}^{\dag}_{-\bm{k},\downarrow,\rm A},\hat{c}^{\dag}_{-\bm{k},\downarrow,\rm B},\hat{c}^{\dag}_{-\bm{k},\downarrow,\rm C}
    \big)^{\!\top}
\end{equation}
and
\begin{equation}\label{eq:BdG_Hamiltonian}
    H_{\rm BdG}(\bm{k}) = \left(\begin{matrix}
        H_0(\bm{k})-\mu\mathbbm{1} & \Delta(\bm{k})\\[1pt]
        \Delta^{\dag}(\bm{k}) & -H_0^\top(-\bm{k})+\mu\mathbbm{1}
    \end{matrix}\right).
\end{equation}
With eigenvalues $E^{(n)}_{\bm{k}}$ and eigenvectors $\bm{\phi}^{(n)}_{\bm{k}} = (\bm{u}^{(n)}_{\bm{k}}\!,\,\bm{v}^{(n)}_{\bm{k}})^\top$ of the Bogoliubov--de Gennes (BdG) equation $H_{\rm BdG}(\bm{k})\bm{\phi}^{(n)}_{\bm{k}} = E^{(n)}_{\bm{k}}\bm{\phi}^{(n)}_{\bm{k}}$, we obtain
\begin{equation}\label{eq:Delta_self-consistent}
    \Delta_{\alpha\beta}(\bm{k}) = -\sum_{\bm{k}'} V_{\alpha\beta}(\bm{k}-\bm{k}')
    \hspace{-3mm}\sum_{\{n|\,E^{(n)}_{\bm{k}'}>0\}} \hspace{-3mm}u^{(n)}_{\bm{k}'\!,\alpha} 
    v^{(n)*}_{\bm{k}'\!,\beta} \tanh\!\bigg(\!\frac{E^{(n)}_{\bm{k}'}}{2T}\!\bigg).
\end{equation}
Note that the BdG Hamiltonian $H_{\rm BdG}(\bm{k})$ possesses particle--hole symmetry, so its eigenvalues always appear in positive and negative pairs.
By solving the BdG equation and Eq.~(\ref{eq:Delta_self-consistent}) self-consistently starting from an initial guess of $\Delta(\bm{k})$, a nonzero solution for $\Delta(\bm{k})$ is obtained when the system is in a superconducting state.

\begin{figure*}[t]
\begin{center}
    \includegraphics[width=\linewidth]{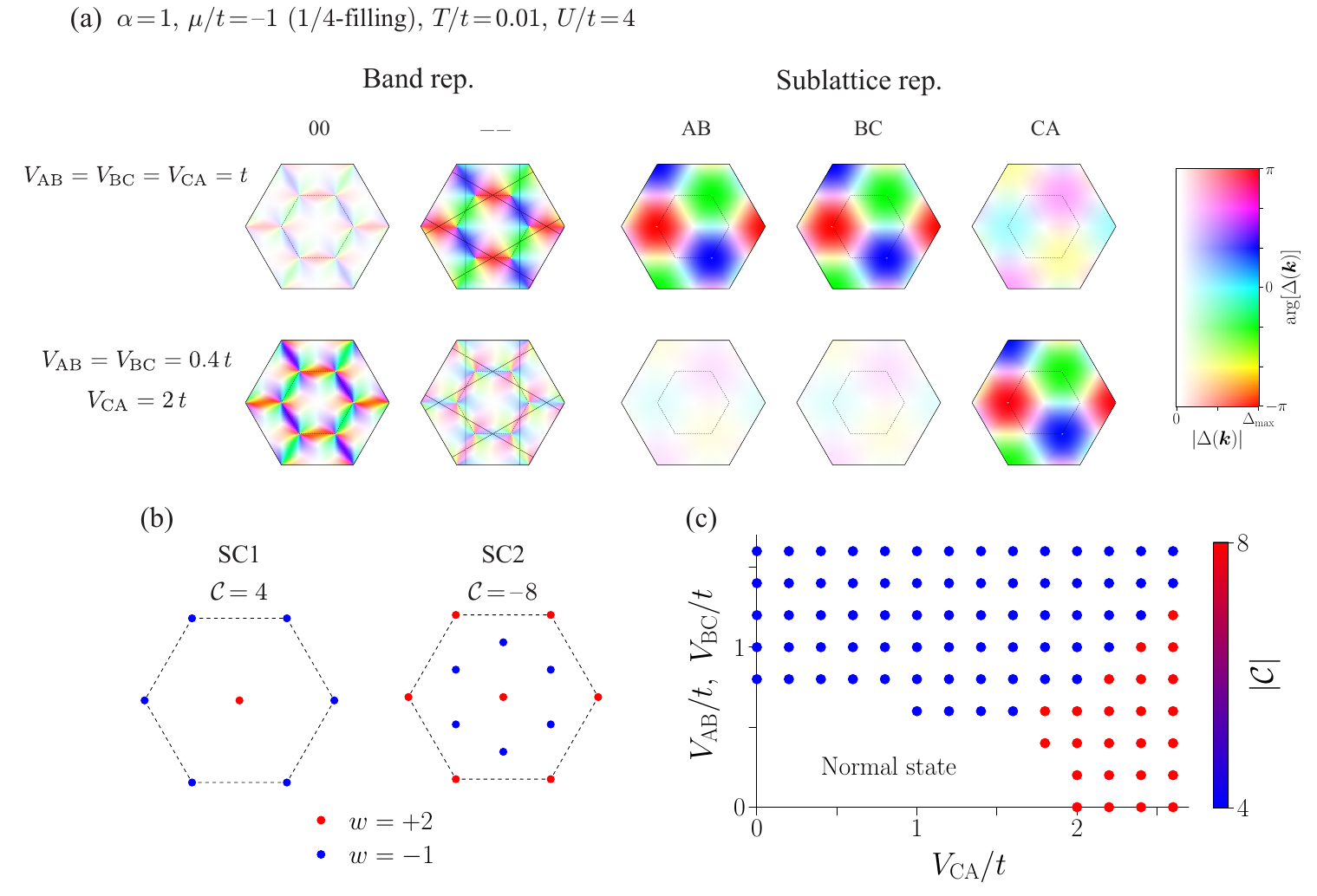}
    \caption{
    (a)~Band and sublattice representations of the pairing potentials $\Delta(\bm{k})$ obtained by self-consistently solving the BdG equation  using an extended Hubbard model with nearest-neighbor attractive interactions.
    Two representative parameter sets of the attractions ($V_{\rm AB}=V_{\rm BC}=V_{\rm CA}$ and $V_{\rm CA}\gg V_{\rm AB}=V_{\rm BC}$) are shown.
    $\Delta(\bm{k})$ is normalized with its maximum amplitude.
    The region enclosed by the dashed lines denotes the first Brillouin zone (BZ), and the black solid contour in the ($--$) panels indicates the Fermi surface of the single-particle band. 
    (b)~Distribution of the nodal points and the corresponding winding numbers of the $(--)$ component of the $d+id'$-wave pairing potential $\Delta_{--}(\bm{k})$ in the first BZ.
    As the interaction strengths between sublattices are varied, the total Chern number $\mathcal{C}$ of the occupied BdG bands switches between $\mathcal{C}=4$ (SC1 phase) and $\mathcal{C}=-8$ (SC2 phase).
    (c)~Superconducting phase diagram for $V_{\rm CA}$ versus $V_{\rm AB}=V_{\rm BC}$.
    When $V_{\rm CA}$ is much larger than $V_{\rm AB}$ and $V_{\rm BC}$, the system is in the SC2 phase.
    The band-representation component with the largest amplitude of $\Delta$ is $(--)$ for the SC1 phase and $(00)$ for the SC2 phase.
    The total Chern numbers were calculated using Fukui--Hatsugai--Suzuki's method~\cite{Fukui_2005}.
    } 
    \label{fig:BdG_plot}
\end{center}
\end{figure*}

\begin{figure}[t]
\begin{center}
    \includegraphics[width=\columnwidth]{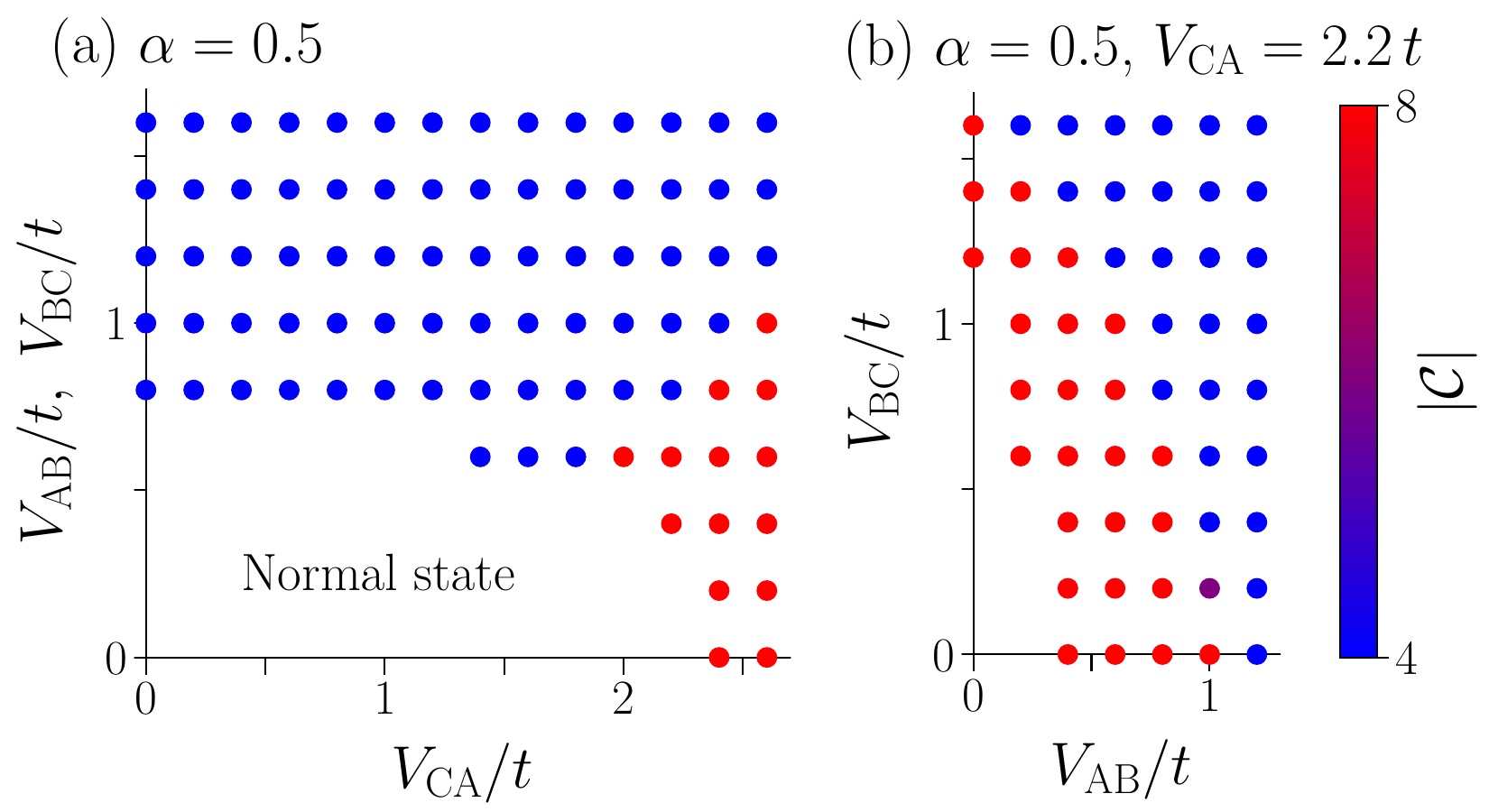}
    \caption{
    Superconducting phase diagrams for (a)~$V_{\rm CA}$ versus $V_{\rm AB}=V_{\rm BC}$ and (b)~$V_{\rm AB}$ versus $V_{\rm BC}$ with $V_{\rm CA}$ fixed, calculated at $\alpha=0.5$.
    All other parameters are the same as in Fig.~\ref{fig:BdG_plot}.
    } 
    \label{fig:phase_diagram_alpha=0.5}
\end{center}
\end{figure}

In Fig.~\ref{fig:BdG_plot}, we present the results of the self-consistent calculation of the pairing potential $\Delta(\bm{k})$ for $\alpha = 1$ (dice lattice).
The chemical potential, temperature, and on-site Coulomb interaction are set to $\mu/t = -1$ (quarter-filling), $T/t = 0.01$, and $U/t = 4$, respectively.
The Fermi level lies at the van Hove singularity (vHS).
We note that the results presented below remain qualitatively unchanged for chemical potentials near quarter filling, apart from differences in the transition temperature  ($T_{\rm c}$).
Figure~\ref{fig:BdG_plot}(a) shows the pairing potentials obtained for two representative parameter sets of the nearest-neighbor attractive interactions $V_{\rm AB}$, $V_{\rm BC}$, and $V_{\rm CA}$.
In the case of $V_{\rm AB}=V_{\rm BC}=V_{\rm CA}$, the AB and BC components of $\Delta(\bm{k})$ have larger amplitudes, whereas for $V_{\rm CA}\gg V_{\rm AB},V_{\rm BC}$, the CA component dominates.
Interestingly, when transformed into the band representation based on Eq.~(\ref{eq:orb2band}), these two parameter sets yield distinct phase structures.
For the ($--$) component, the phase winds once around the $\textrm{K}$($\textrm{K}'$) point for $V_{\rm AB}=V_{\rm BC}=V_{\rm CA}$, whereas it winds twice for $V_{\rm CA}\gg V_{\rm AB},V_{\rm BC}$.
Figure~\ref{fig:BdG_plot}(b) schematically illustrates these features. 
Indeed, these two states belong to different topological phases, characterized by the distinct total Chern numbers, defined as
\begin{equation}
    \mathcal{C} = \frac{1}{2\pi i} \sum_{n\in{\rm occ}} \int_{\rm BZ}\! d^2\bm{k} 
    \bigg(\frac{\partial A_y^{(n)}}{\partial k_x} - \frac{\partial A_x^{(n)}}{\partial k_y}\bigg),
\end{equation}
where the sum runs over occupied BdG bands and $\bm{A}^{(n)} = \bra{\phi_n(\bm{k})}\nabla_{\!\bm{k}}\ket{\phi_n(\bm{k})}$ is the Berry connection of the $n$th BdG band, with $\ket{\phi_n(\bm{k})}$ denoting the corresponding eigenstate of the BdG Hamiltonian~(\ref{eq:BdG_Hamiltonian}).
When the phase winding number of $\Delta_{--}(\bm{k})$ around the $\textrm{K}$($\textrm{K}'$) point is $w = -1$, the Chern number is $\mathcal{C} = 4$ (SC1 phase), whereas for $w = 2$, the Chern number is $\mathcal{C} = -8$ (SC2 phase).
Here, the Chern number is given counting the spin degree of freedom.
The sign of the Chern number reverses when the sign of the chemical potential $\mu$ is changed.
In Fig.~\ref{fig:BdG_plot}(c), we present the superconducting phase diagram as a function of $V_{\rm AB}=V_{\rm BC}$ (vertical axis) and $V_{\rm CA}$ (horizontal axis).
As long as the on-site interaction is repulsive ($U > 0$), both the transition temperature and the phase boundary are almost independent of~$U$.

The Chern number $|\mathcal{C}|=4$ is the same as that obtained for spinless graphene assuming $d+id'$-wave pairing potential, while the Chern number $|\mathcal{C}|=8$ can be realized in appropriately tuned AB-stacked bilayer graphene assuming a $d+id'$-wave pairing potential with a phase difference of $\pi$ between the layers~\cite{Crepieux_2023}.
The possibility of $d+id'$-wave superconductivity with $|\mathcal{C}|=8$ has  also been pointed out in multiorbital or spin--orbit coupled triangular lattices~\cite{C-Lu_2018,Matthew_2024}.
Our results are interesting in that a chiral superconducting state with $|\mathcal{C}|=8$ is realized in a system where sites are connected in a rather simple way.

In Fig.~\ref{fig:phase_diagram_alpha=0.5}, we present the superconducting phase diagram for $\alpha = 0.5$.
All other parameters are the same as those in Fig.~\ref{fig:BdG_plot}.
Figure~\ref{fig:phase_diagram_alpha=0.5}(a) shows, as in Fig.~\ref{fig:BdG_plot}(c), the superconducting phase diagram as a function of $V_{\rm AB}=V_{\rm BC}$ and $V_{\rm CA}$. 
As in the case of $\alpha = 1$, the SC2 phase appears when $V_{\rm CA}\gg V_{\rm AB},V_{\rm BC}$.
Compared with the case of $\alpha = 1$, the boundary between the SC2 and normal phases shifts toward larger $V_{\rm CA}$ at the same temperature, indicating that $T_{\rm c}$ of the SC2 phase decreases as $\alpha$ is reduced.
In contrast, $T_{\rm c}$ of the SC1 phase remains almost unchanged from that for $\alpha = 1$.
For $\alpha=0.5$, the rim sites A and C are no longer equivalent, and it is therefore useful to consider the case where $V_{\rm AB} \neq V_{\rm BC}$.
Figure~\ref{fig:phase_diagram_alpha=0.5}(b) presents the superconducting phase diagram obtained with $V_{\rm CA}$ fixed. 
The phase boundary is asymmetric with respect to $V_{\rm AB}$ and $V_{\rm BC}$, which can be attributed to the larger hopping amplitude $t_{\rm AB}$ compared with $t_{\rm BC}$.

\subsection{Spin-fluctuation-mediated $d+id'$-wave superconductivity in the repulsive Hubbard model}

Within the mean-field approximation, assuming an off-site attractive interaction yields a $d+id'$-wave superconducting ground state.
Here, we explore the possibility that chiral superconductivity can emerge from purely repulsive interactions, where the pairing is mediated by spin fluctuations.
We consider the on-site Coulomb repulsion
\begin{equation}
    \hat{H}_{\rm int} = U\sum_{i} \,\hat{n}_{i,\uparrow} \hat{n}_{i,\downarrow}
\end{equation}
as the two-body interaction term.
To investigate spin-fluctuation-mediated superconductivity in the Hubbard model, we calculate the self-energy within the FLEX approximation~\cite{Bickers_1989,Dahm_1995} and obtain the renormalized Green's function $G(\bm{k},i\omega_n)$.
Using this, we solve the linearized Eliashberg equation
\begin{align}\label{eq:Eliash_eq}
    &\lambda \Delta_{ll'}(\bm{k},i\omega_n)\notag\\
    &= -\frac{T}{N}\!\!\sum_{\bm{k}'\!,n'\!,m_i}\!\!
    \Gamma_{lm_1,m_4l'}(\bm{k}\!-\!\bm{k}', i\omega_n\!-\!i\omega_{n'}) 
    G_{m_1m_2}(\bm{k}',i\omega_{n'})\notag\\
    &\hspace{40pt}\times \Delta_{m_2m_3}(\bm{k}',i\omega_{n'})G_{m_4m_3}(-\bm{k}',-i\omega_{n'})
\end{align}
to obtain the eigenvalue $\lambda$ and the gap function $\Delta$. 
Here, $T$, $N$, and $\Gamma$ denote the temperature, the number of the $\bm{k}$-points, and the pairing interaction, respectively.
We regard $\lambda$, calculated at a fixed temperature, as a quantity representing how high the superconducting critical temperature $T_{\mathrm{c}}$ of the system is.
Unless otherwise specified, we use $U/t=4$, $T/t=0.002$, $4096\times2$ Matsubara frequencies, and $48\times48$ $\bm{k}$-mesh.
See Appendix~\ref{app:detailed_analysis} for results with different parameters.
In this study, we assume a spin-singlet pairing interaction:
\begin{align}\label{eq:pairing_interaction_singlet}
    \Gamma(\bm{q},i\Omega_m)
    &=\frac32U^{\rm s}\chi^{\rm s}(\bm{q},i\Omega_m)U^{\rm s} 
    - \frac12U^{\rm c}\chi^{\rm c}(\bm{q},i\Omega_m)U^{\rm c} \notag\\
    &\quad+ \frac12\left(U^{\rm s}\!+\!U^{\rm c}\right),
\end{align}
where the spin and charge susceptibility ($\chi^{\rm s}$ and $\chi^{\rm c}$) are expressed as
\begin{equation}
\begin{aligned}
    \chi^{\rm s}(\bm{q},i\Omega_m) &= \chi^0(\bm{q},i\Omega_m)\left[
    \mathbbm{1}-U^{\rm s}\,\chi^0(\bm{q},i\Omega_m)
    \right]^{-1},\\[1pt]
    \chi^{\rm c}(\bm{q},i\Omega_m) &= \chi^0(\bm{q},i\Omega_m)\left[
    \mathbbm{1}+U^{\rm c}\,\chi^0(\bm{q},i\Omega_m)
    \right]^{-1},
\end{aligned}
\end{equation}
in terms of the irreducible susceptibility,
\begin{align}
    &\chi^0_{l_1l_2,l_3l_4}(\bm{q},i\Omega_m) \notag\\
    &= -\frac{T}{N} \sum_{\bm{k},n} 
    G_{l_1l_3}(\bm{k},i\omega_{n})
    G_{l_4l_2}(\bm{k}\!+\!\bm{q},i\omega_{n}\!+\!i\Omega_m).
\end{align}
Since we only consider the on-site Coulomb interaction, $U^{\rm s}_{l_1l_2,l_3l_4}=U^{\rm c}_{l_1l_2,l_3l_4}=U\delta_{l_1,l_2}\delta_{l_1,l_3}\delta_{l_1,l_4}$ holds.
We present the eigenvalue $\lambda$ when assuming the spin-triplet pairing interaction in Appendix~\ref{app:detailed_analysis}.
Note that for almost all parameter sets of interest, we have $\lambda_{\rm singlet} > \lambda_{\rm triplet}$.

\begin{figure}[t]
\begin{center}
    \includegraphics[width=\columnwidth]{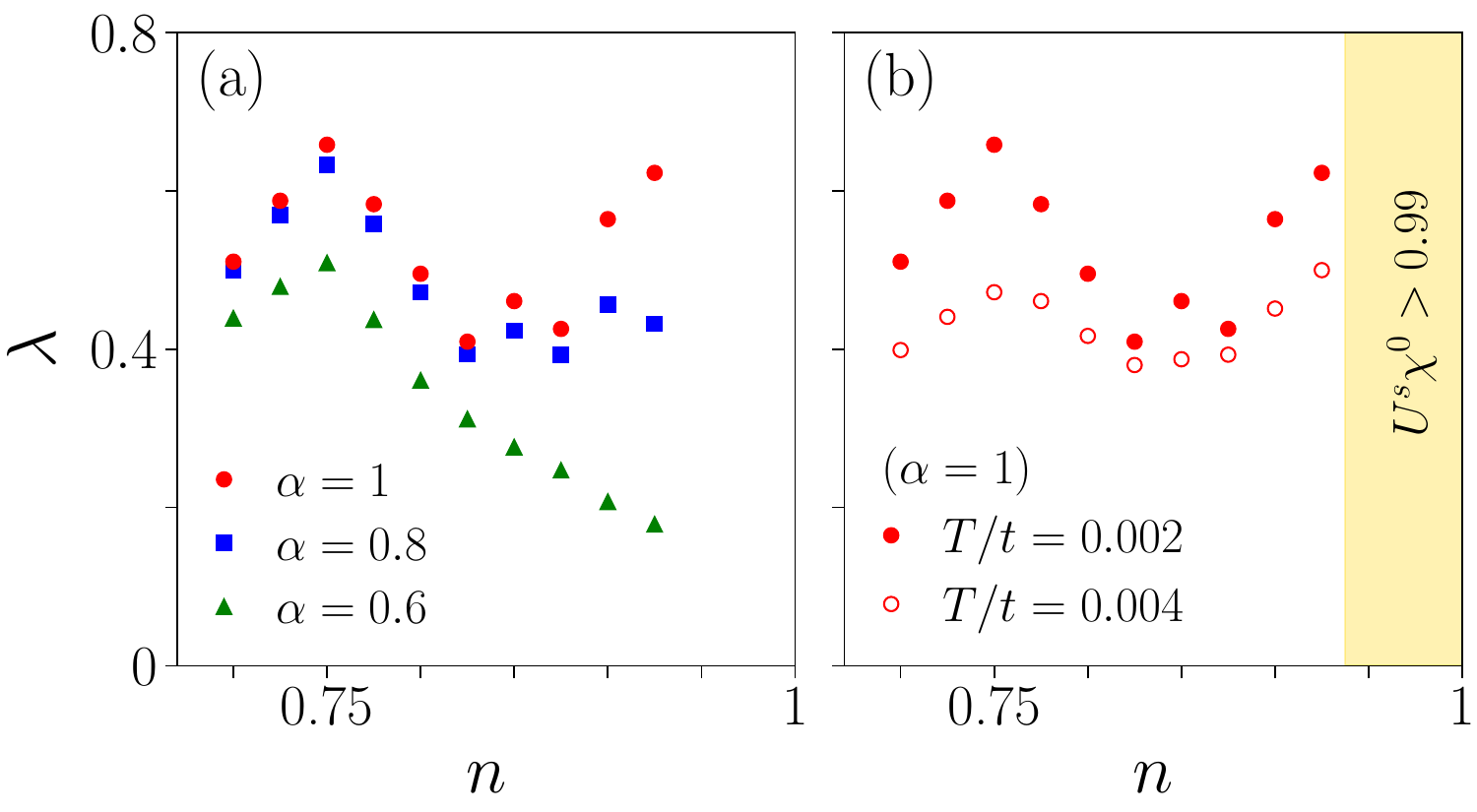}
    \caption{
    Band filling $n$ dependence of the eigenvalue $\lambda$ of the linearized Eliashberg equation~(\ref{eq:Eliash_eq}) for (a)~various values of $\alpha$ at $T/t=0.002$ and (b)~various values of $T$ at $\alpha=1$.
    Band fillings $n=0.75$ and $n=1$ correspond to the $1/4$- and $1/3$-filling, respectively.
    The gap functions $\Delta(\bm{k})$ are the $d$-wave for all $n$.
    The maximum eigenvalue of $U^{\rm s}\chi^0(\bm{q})$ (Stoner factor) approaches to $1$ at $\bm{q}=(0,0)$ near the $1/3$-filling.} 
    \label{fig:lambda}
\end{center}
\end{figure}

In Fig.~\ref{fig:lambda}, we show the band filling $n$ dependence of eigenvalues $\lambda$ of the linearized Eliashberg equation.
The band filling $n$ is defined as the number of electrons per spin per unit cell, e.g., $n=3$, $n=1$, and $n=0.75$ correspond to full-filling, $1/3$-filling, and $1/4$-filling, respectively.
For all the parameter sets we examined, the $d$-wave gap symmetry is obtained.
As shown in Fig.~\ref{fig:lambda}(a), $\lambda$ takes its maximum at $\alpha=1$.
For all values of $\alpha$, $\lambda$ shows a peak structure near the $1/4$-filling, at the vHS.
For $\alpha=1$, $\lambda$ increases toward the $1/3$-filling.
This behavior may be attributed to a trade-off between the reduction of the density of states (DOS) as the Fermi level moves away from the vHS and the enhancement of spin fluctuations as the Fermi level approaches the flat band toward the $1/3$-filling.
The spin fluctuations that can favor $d$-wave superconductivity will be discussed later.

In Fig.~\ref{fig:lambda}(b), we show how $\lambda$ changes with temperature $T$.
From its temperature dependence, the superconducting transition temperature $T_{\mathrm{c}}$ is expected to take the maximum value at $n=0.75$ ($1/4$-filling).
For $n=0.75$, the transition temperature at which $\lambda$ reaches $1$ is estimated as $T_{\mathrm{c}}\approx7.5\times10^{-4}\,t$ at $n=0.75$ by extrapolating $\lambda$ calculated at $T/t=0.004$, $0.003$, $0.002$, and $0.001$ (see Appendix~\ref{app:detailed_analysis} for details).
Near $n=1$ ($1/3$-filling), the maximum eigenvalue of $U^{\rm s}\chi^0(\bm{q})$ (Stoner factor) approaches $1$, suggesting that magnetic order tends to be favored over superconductivity.
If the Fermi level lies within the flat band ($n>1$), correlation effects become more intricate and require a rigorous treatment of interactions; therefore, we do not consider this case.
As a reference, a study of the one-dimensional diamond chain based on exact diagonalization and density matrix renormalization group (DMRG) calculations suggest that superconductivity is suppressed when the Fermi level lies within the flat band~\cite{Kobayashi_2016}. This regime will be discussed again in Sec.~\ref{sec:flat_band}.

\begin{figure*}[p]
\begin{center}
    \includegraphics[width=\linewidth]{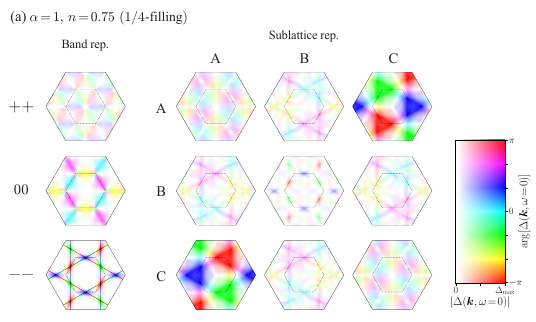}
    \includegraphics[width=\linewidth]{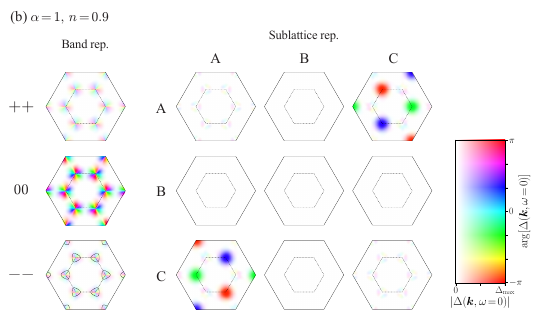}
    \caption{
    Band and sublattice representations of the gap functions $\Delta(\bm{k},\omega\!=\!0)$ obtained at $\alpha=1$ and band fillings (a)~$n=0.75$ and (b)~$n=0.9$.
    The hue represents the phase of the gap function, while the brightness indicates its amplitude.
    The black solid contour in the ($--$) panels indicates the Fermi surface of the single-particle band.
    For all the parameters we examined, the phase winding number around the $\textrm{K}$($\textrm{K}'$) point in the band representation is $|w|=2$, which is consistent with the SC2 phase ($|\mathcal{C}|=8$) obtained from the mean-field analysis of the extended Hubbard model.
    } 
    \label{fig:gap}
\end{center}
\end{figure*}

In Fig.~\ref{fig:gap}, we show the gap functions obtained at two representative parameters, $n=0.75$ and $n=0.9$, for $\alpha=1$.
The gap functions were first calculated on the Matsubara frequency axis and then analytically continued to the real-frequency axis using the Pad\'{e} approximation to obtain their $\omega=0$ components.
In a hexagonal lattice structure (point group $D_{6h}$), the linearized Eliashberg equation gives the same $T_{\mathrm{c}}$ for any basis function belonging to a $E_{2g}$ irreducible representation~\cite{Black-Schaffer_2014_review}.
Therefore, the gap function obtained from the linearized Eliashberg equation is expressed as a linear combination of the $d_{x^2-y^2}$ and $d_{xy}$ components.
Below $T_{\mathrm{c}}$, higher-order terms lift this degeneracy and stabilize a specific superconducting state.
In general, the chiral $d+id'$-wave state, which opens a full gap in the quasiparticle spectrum, is energetically favored~\cite{Black-Schaffer_2014_review}.
Accordingly, in Fig.~\ref{fig:gap}, we orthogonalize the two degenerate gap functions obtained from the linearized Eliashberg equation and present them as the $d+id'$-wave state.

\begin{figure*}[t]
\begin{center}
    \includegraphics[width=\linewidth]{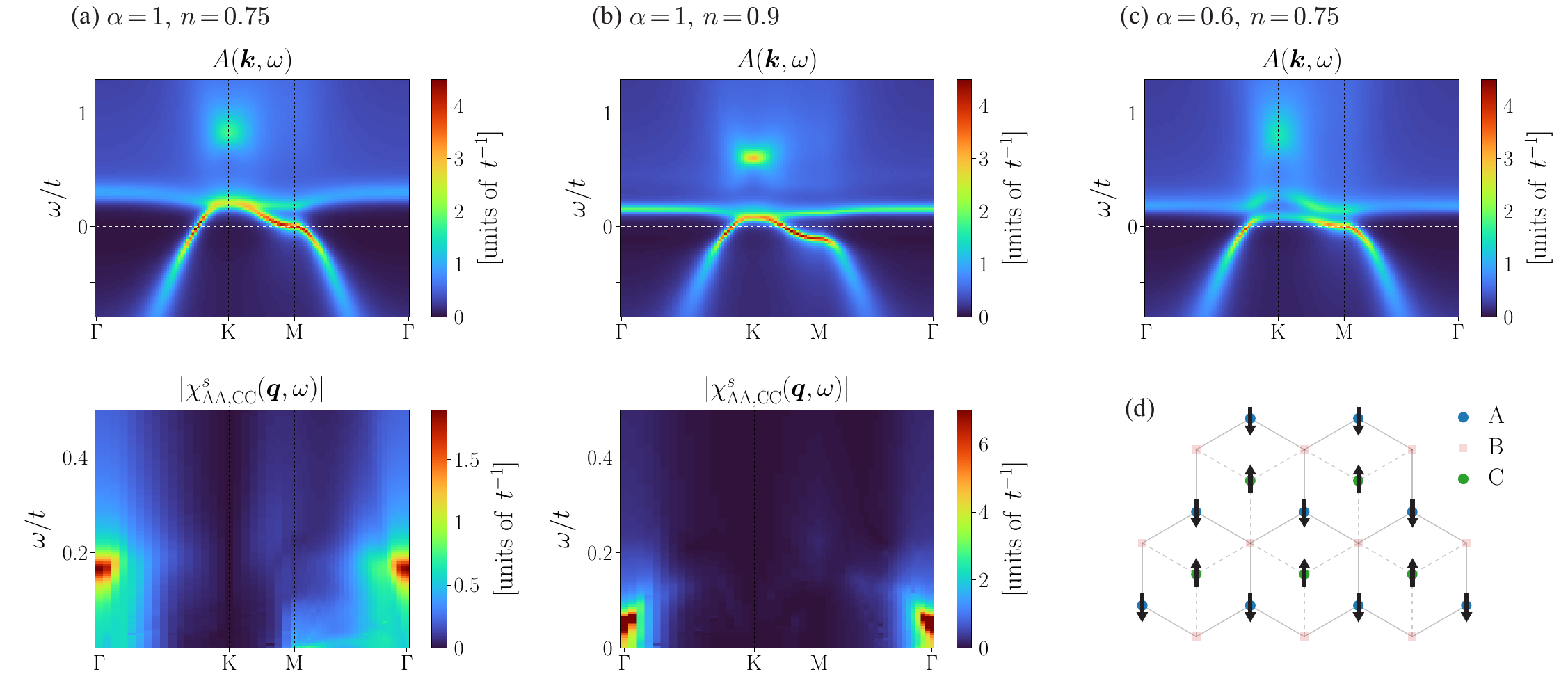}
    \caption{
    Spectral functions $A(\bm{k},\omega)$ (top panels) and the AC components of the dynamical spin susceptibilities $\chi^{\rm s}_{\rm AA,CC}(\bm{q},\omega)$ (bottom panels) for $\alpha=1$, and band fillings (a)~$n=0.75$ and (b)~$n=0.9$. 
    The spectral function for $\alpha=0.6$ and $n=0.75$ is shown in~(c) for comparison.
    In the dynamical spin susceptibility, the largest spectral weight appears at finite energy near $\bm{q}=(0,0)$.
    (d)~Schematic illustration of the real-space spin configuration corresponding to the $\bm{q}=\bm{0}$ spin fluctuation between rim sites.
    Note that each arrow does not represent a single electron.
    } 
    \label{fig:spec_suscep_schematic}
\end{center}
\end{figure*}

Interestingly, in the band representation, both gap functions shown in Fig~\ref{fig:gap}(a) and (b) exhibit a phase winding number of $|w|=2$ around the $\textrm{K}$($\textrm{K}'$) point.
This feature is the same as that of the superconducting state with a Chern number $|\mathcal{C}|=8$ obtained from the mean-field analysis of the extended Hubbard model, indicating that the system is in the SC2 phase at low temperatures.
In the sublattice representation as well, the AC component exhibits the largest amplitude.
This characteristic becomes more pronounced as the band filling approaches $n=1$ ($1/3$-filling).
The result that the superconducting state obtained from the FLEX approximation for the pure (on-site repulsive) Hubbard model qualitatively agrees with that obtained from the mean-field analysis of the extended Hubbard model, where an attractive interaction is assumed between the rim (A and C) sites.
This indicates that the off-site attractive Hubbard model ($V_{\rm AC}\gg V_{\rm AB},V_{\rm BC}$) studied in Sec.~\ref{sec:mean-field} can be considered as an effective model that captures the essential physics of superconductivity near the quarter-filling in the repulsive Hubbard model.

\subsection{Pairing mechanism: Finite-energy spin fluctuations between the rim sites}

To identify the pairing mechanism of the $d+id'$-wave superconductivity in the Hubbard model on the $\alpha$--$\mathcal{T}_3$ lattice, we examine the properties of the renormalized Green's function.
In Figs.~\ref{fig:spec_suscep_schematic}(a) and \ref{fig:spec_suscep_schematic}(b), we present the spectral functions defined as $A(\bm{k},\omega)={\rm Tr}[-\tfrac{1}{\pi}{\rm Im}\,G(\bm{k},\omega)]$ and the AC component of the dynamical spin susceptibility $\chi^{\rm s}_{\rm AA,CC}(\bm{q},\omega)$ calculated at $\alpha=1$.
The parameters used in Figs.~\ref{fig:spec_suscep_schematic}(a) and \ref{fig:spec_suscep_schematic}(b) are the same as those in Figs.~\ref{fig:gap}(a) and \ref{fig:gap}(b).
These quantities are calculated on the Matsubara frequency axis and then analytically continued to the real-frequency axis using the Pad\'{e} approximation.
The full $3\times3$ components of the dynamical spin susceptibility for $n=0.75$ are presented in Appendix~\ref{app:chi_S_dynamical}.

The spectral function indicates that, due to electron correlations, a flat band approaches the Fermi level compared with the single-particle band shown in Fig.~\ref{fig:alpha-T3_lattice_and_band}(b).
This can be qualitatively explained by a model with a potential offset between the hub and rim sites ($\varepsilon_{\rm B}>\varepsilon_{\rm A}=\varepsilon_{\rm C}$)~\cite{Piechon_2015}, corresponding to the Hartree approximation.
In this case, the degeneracy between the upper and lower bands at the $\textrm{K}$($\textrm{K}'$) point is lifted, while the degeneracy between the lower and flat bands remains~\footnote{While on-site potential terms preserve the flat band~\cite{Piechon_2015}, electron correlation effects, which give a self-energy corresponding to long-range hopping, can break it.}.
We also present the spectral function calculated at $\alpha=0.6$ in Fig.~\ref{fig:spec_suscep_schematic}(c).
Since the A and C sites are no longer equivalent, the behavior of $A(\bm{k},\omega)$ can be qualitatively understood by taking into account a potential offset between these sites as well ($\varepsilon_{\rm A},\varepsilon_{\rm B}>\varepsilon_{\rm C}$).

Next, we examine the AC component of the spin fluctuations, which is expected to be important based on the gap function and mean-field analysis.
In the static spin susceptibility ($\omega = 0$), the maximum intensity appears at the M point for $n = 0.75$ [see Fig.~\ref{fig:spec_suscep_schematic}(a)] and at the $\Gamma$ point for $n = 0.9$ [see Fig.~\ref{fig:spec_suscep_schematic}(b)], seemingly without any common feature.
Despite the difference in the static component, focusing on $\omega > 0$, we find that in both cases the strongest intensity occurs at $\bm{q} = \bm{0}$ with finite energy.
The energy at which the maximum intensity appears is higher for $n = 0.75$ and approaches $\omega = 0$ as the band filling approaches $n=1$ ($1/3$-filling).
This energy scale corresponds to the relative position between the flat band and the Fermi level, implying that the flat band drives the finite-energy spin fluctuations.
The qualitative agreement in the dynamical component is consistent with the similar behavior of the gap functions in Figs.~\ref{fig:gap}(a) and \ref{fig:gap}(b).

We note that the superconducting mechanism mediated by finite-energy spin fluctuations has been discussed in the context of iron-based superconductors~\cite{Wang_2011,Bang_2014,Bang_2016,X-Chen_2015,Linscheid_2016} with some experimental support~\cite{Miao_2015,Charnukha_2015,Nishioka_2021,Kouchi_2022}.
Subsequently, it has been recognized as a generic mechanism that enhances $s_{\pm}$-wave superconductivity in systems with an incipient band~\cite{Mishra_2016,Nakata_2017,Ogura_2018,Matsumoto_2018,Matsumoto_2020,Maier_2019,Kato_2020,Aida_2024}, which may account for the superconductivity observed in, e.g., the ladder-type cuprate Pr$_2$Ba$_4$Cu$_7$O$_{15-\delta}$~\cite{T-Nakano_2007,Yagi_2024} and Ruddlesden--Popper nickelates~\cite{Nakata_2017,Sakakibara_2024a,Sakakibara_2024b}.

To intuitively understand how the spin fluctuation at $\bm{q} = \bm{0}$ acts as the pairing glue for $d$-wave superconductivity, we review the triangular lattice with dilute band filling and the nearly half-filled (bipartite) honeycomb lattice~\cite{Kuroki_2001,Kuroki_2010}.
In a triangular lattice with one site per unit cell, the $\bm{q} = \bm{0}$ fluctuation corresponds to ferromagnetic correlations and favors $f$-wave pairing, whereas in a honeycomb lattice with two sites per unit cell, it corresponds to antiferromagnetic correlations between nearest-neighbor sites and favors $d$-wave pairing~\footnote{When the potential offset is sufficiently large, the nearly half-filled honeycomb lattice continuously connects to the triangular lattice with dilute band filling, where $f$-wave pairing is favored~\cite{Kuroki_2001}}.
In Fig.~\ref{fig:spec_suscep_schematic}(d), we present the real-space spin configuration that is naturally favored by the $\bm{q} = \bm{0}$ fluctuation of the AC component.
Note that each arrow in the figure does not correspond to a single electron.
As detailed in Appendix~\ref{app:chi_S_dynamical}, all other components of $\chi^{\rm s}(\bm{q},\omega)$ also support this configuration; specifically, the diagonal AA and CC components exhibit the largest spectral weight at $\bm{q} = \bm{0}$ at finite energy, while the BB, AB, and BC components show weaker intensity.
Focusing on the rim sites (A and C), although there is no direct connection, their real-space geometry is equivalent to that of the honeycomb lattice.
Therefore, it is reasonable to associate the pairing glue of the spin-singlet superconductivity in the $\alpha$--$\mathcal{T}_3$ lattice with the spin fluctuation at $\bm{q} = \bm{0}$ between the rim sites.
While such an analogy exists, it is noteworthy that in the honeycomb and triangular lattices, the $\bm{q} = \bm{0}$ spin fluctuation originates purely from the nesting of small Fermi surfaces, whereas in the $\alpha$--$\mathcal{T}_3$ lattice, the dominant contribution arises from the flat band near the Fermi level.

Interactions between the rim sites might also be interpreted as an analog of the superexchange interaction in cuprates~\cite{Anderson_1950,*Anderson_1959,Emery_1987}.
In cuprates, an antiferromagnetic interaction arises between electrons on Cu sites through a fourth-order perturbation process involving intermediate states in which either an O site or a Cu site is doubly occupied.
Since the difference in electron occupancy between the hub and rim sites effectively raises the energy level of the hub site compared to the rim sites ($\varepsilon_{\rm B}>\varepsilon_{\rm A}=\varepsilon_{\rm C}$), we may associate the rim sites with the Cu sites and the hub sites with the O sites.
From a given rim site (denoted as the A site for simplicity), there are 12 rim sites that can be reached via two hopping processes through a hub site (six A sites and six C sites). Among these, there are twice as many paths to the three nearest-neighbor C sites as to the other sites, suggesting that stronger magnetic interactions are induced along these routes.
Consequently, although there are frustrations among equivalent rim sites in this process, spin configurations involving spin flips between inequivalent rim sites may be energetically favored overall.
Although a direct comparison with the cuprates is not possible due to the difference in electron density and the presence of frustration, this analogy may still provide an intuitive understanding of the magnetic interactions between the rim sites.

\subsection{Extended discussion}

\subsubsection{Robustness of superconductivity against band-structure modifications}

\begin{figure}[t]
\begin{center}
    \includegraphics[width=\columnwidth]{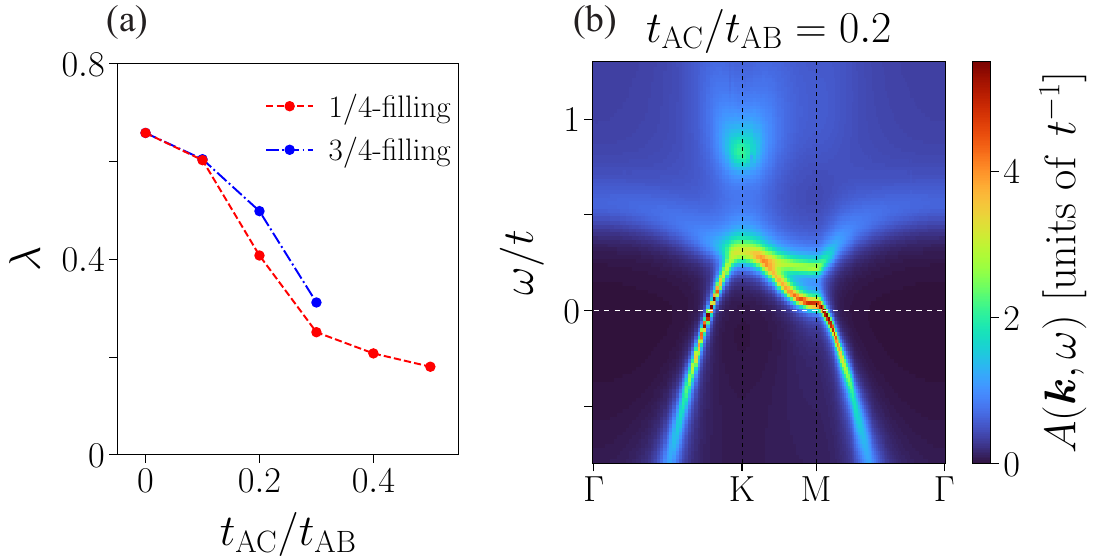}
    \caption{
    (a)~$t_{\rm AC}$ dependence of the eigenvalue $\lambda$ of the linearized Eliashberg equation for $t_{\rm AB}=t_{\rm BC}$ ($\alpha=1$). Because introducing $t_{\rm AC}$ breaks the particle--hole symmetry of the band dispersion, results at both $1/4$- and $3/4$-filling ($n=0.75$ and $n=2.25$) are presented.
    (b)~Spectral function $A(\bm{k},\omega)$ for $t_{\rm AC}/t_{\rm AB}=0.2$ at $1/4$-filling.
    }
    \label{fig:t_AC}
\end{center}
\end{figure}

We examine the robustness of the superconductivity obtained in this work against modifications to the tight-binding band structure, in particular against perturbations that break the flat band. For this purpose, we introduce an additional hopping $t_{\rm AC}$ between rim sites, which breaks the bipartite structure of the Hamiltonian and consequently makes the flat band dispersive in the noninteracting system.
In Fig.~\ref{fig:t_AC}(a), we show the $t_{\rm AC}$ dependence of the eigenvalue $\lambda$ of the linearized Eliashberg equation for  $\alpha=1$. The ratio $t_{\rm AC}/t_{\rm AB}=1$ corresponds to the triangular lattice. Since the particle--hole symmetry of the band dispersion is broken, $\lambda$ takes different values at $1/4$- and $3/4$-filling.
The $d$-wave solution of the linearized Eliashberg equation gives the largest eigenvalue within the parameter range shown in Fig.~\ref{fig:t_AC}(a).
We note that the eigenvalues of the $d$-wave solution remain degenerate irrespective of $t_{\rm AC}$.
As shown in Fig.~\ref{fig:t_AC}(a), $\lambda$ decreases continuously as $t_{\rm AC}/t_{\rm AB}$ increases, indicating that the superconductivity is moderately robust against modifications of the band structure. Superconductivity mediated by finite-energy spin fluctuations generally exhibits weak sensitivity to the detailed band structure near the Fermi level.
The spectral function for $t_{\rm AC}/t_{\rm AB}=0.2$ at quarter-filling is shown in Fig.~\ref{fig:t_AC}(b) for reference.
The variation of $\lambda$ with $t_{\rm AC}$ may be attributed to multiple factors, such as the dispersion of the flat band and changes in the DOS. Identifying which contribution is dominant is beyond the scope of this paper.

\subsubsection{When the Fermi level intersects the flat band\label{sec:flat_band}}

Finally, we discuss the case where the band filling exceeds $1/3$-filling and approaches half-filling.
In this regime, the Fermi level lies within the flat band, making a ferromagnetic (ferrimagnetic) ground state likely~\cite{Lieb_1989,Mielke_1991a,*Mielke_1991b,Tasaki_1992,*Tasaki_1994,Tanaka_2003}.
Near half-filling, magnetic interactions between the hub and rim sites (A--B and B--C) become dominant and compete with the spin configuration shown in Fig.~\ref{fig:spec_suscep_schematic}(d), thereby suppressing the $d+id'$-wave superconductivity in the SC2 phase.
The nearest-neighbor magnetic interaction in systems with the dice lattice network naturally induces ferrimagnetism, which has been confirmed by exact diagonalization and DMRG calculations~\cite{H-Nakano_2017,Soni_2020}.
As a possible scenario for superconductivity in the flat band, spin-triplet superconductivity induced by ferromagnetic fluctuations has been discussed in the context of quantum geometry~\cite{Peotta_2015,Kitamura_2024}.
Another is that magnetic interactions between the hub and rim sites might promote the $d+id'$-wave superconductivity in the SC1 phase.
In addition, it is known that introducing Rashba spin--orbit coupling into the dice lattice ($\alpha=1$) leads to the emergence of a topological flat band~\cite{F-Wang_2011}. 
Recent work has shown that attractive interaction induces topological superconductivity with mixed spin-singlet and spin-triplet components within this flat band~\cite{Mohanta_2023}.
In a context distinct from topological superconductivity, a recent study investigated superconductivity in the flat band of an on-site attractive ($U<0$) Hubbard model on the $\alpha$--$\mathcal{T}_3$ lattice using a mean-field approximation, revealing the contribution of the quantum metric to the superfluid weight~\cite{Mojarro_arxiv}.
Excitonic gap generation has also been predicted at half-filling in the dice lattice~\cite{Gorbar_2021}.

\section{Summary}
We investigated anisotropic superconductivity in the nearly quarter-filled $\alpha$--$\mathcal{T}_3$ lattice.
Within the mean-field approximation, we analyze an extended Hubbard model with nearest-neighbor attractive interactions.
A self-consistent solution of the gap equation yields a chiral $d$-wave superconducting ground state that spontaneously breaks time-reversal symmetry.
Varying the attractive interactions among the three sublattices gives rise to two superconducting phases distinguished by their total Chern numbers: $|\mathcal{C}|=4$ (SC1) and $|\mathcal{C}|=8$ (SC2).
As $\alpha$ decreases, the transition temperature $T_c$ of the SC2 phase tends to decrease for the same strength of attraction.

We further investigated spin-fluctuation-mediated superconductivity in the Hubbard model with only on-site repulsive interactions.
From the linearized Eliashberg equation, we obtain a $d$-wave gap function in the range from approximately $1/4$- to below $1/3$-filling.
This behavior is similar to the chiral $d+id'$-wave state in the SC2 phase of the extended attractive Hubbard model, indicating that, interestingly, the SC2 state can be realized in a purely repulsive model. 
The $\bm{q}=\bm{0}$ antiferromagnetic fluctuation between rim sites, with its maximum intensity at a finite energy, likely promotes spin-singlet pairing.
Note that the degeneracy between the eigenvalues for the $d_{x^2-y^2}$- and $d_{xy}$-wave solutions of the linearized Eliashberg equation does not necessarily guarantee that their chiral combination is energetically selected below $T_{\mathrm{c}}$.
Consistency with the gap functions obtained within the mean-field approximation for the off-site attractive Hubbard model---regarded as effectively incorporating spin-fluctuation effects---provides indirect evidence for the stabilization of a chiral state. However, directly demonstrating the energetic stability by self-consistently solving the nonlinear Eliashberg equation below $T_{\mathrm{c}}$ remains a subject for future investigation.

Our results provide a possible pathway for exploring topological superconductivity in generic pseudospin-1 systems with flat bands. 
In particular, recent studies have examined superconducting instabilities near the three-band crossing points, with possible realizations of various nontrivial superconducting states~\cite{YP-Lin_2018,YP-Lin_2020,Mandal_2021a,Mandal_2021b,Sim_2022,Zyuzin_2022}.
Our numerical calculations for different hopping ratios $\alpha$ clearly show that many-body effects depend on both the single-particle band structure and the corresponding wave functions. 
For a comprehensive understanding of superconductivity in these systems, it is important to clarify how the band structure and wave functions contribute to the development of spin fluctuations and other many-body effects.

Finally, we comment on direct experimental probes of $d+id'$-wave superconductivity. 
As an experimental fingerprint of chiral superconductivity, edge currents associated with a finite Chern number and chiral edge modes have been widely discussed~\cite{Kallin_2016_review}. However, edge currents in chiral $d$-wave superconductors can be much smaller than those in chiral $p$-wave superconductors, and may even vanish completely depending on lattice details~\cite{Huang_2014}. 
Recent theoretical studies have suggested that mesoscopic finite-size effects and engineered edge patterning can enhance the edge currents to an experimentally detectable level~\cite{Holmvall_2023c,Holmvall_2025}.
Furthermore, vortex features that can be directly probed by scanning tunneling microscopy (STM) have recently been proposed, potentially providing a means to identify the Chern number~\cite{Holmvall_2023a,*Holmvall_2023b,Cadorim_2024,Yoshida_2025}.

\begin{acknowledgments}
    This work was supported by the JSPS KAKENHI, Grants No.~JP25KJ1758 (M.K.), No.~JP25H01252, and No.~JP24K01333 (K.K.).
    M.K. thanks Fr\'{e}d\'{e}ric Pi\'{e}chon and Masayuki Ochi for helpful discussions.
    M.K. also thanks the Research Fellowship for Young Scientists (Grant No.~JP25KJ1758), the UCA$^\text{JEDI}$ ``Investissements d'Avenir" project, managed by the National Research Agency (ANR) (Grant No.~ANR-15-IDEX-01), and the Program for Leading Graduate Schools: ``Interactive Materials Science Cadet Program" for support.
\end{acknowledgments}

\appendix 

\section{Additional results of the FLEX + linearized Eliashberg equation analyses\label{app:detailed_analysis}}

\begin{figure}[t]
\begin{center}
    \includegraphics[width=\columnwidth]{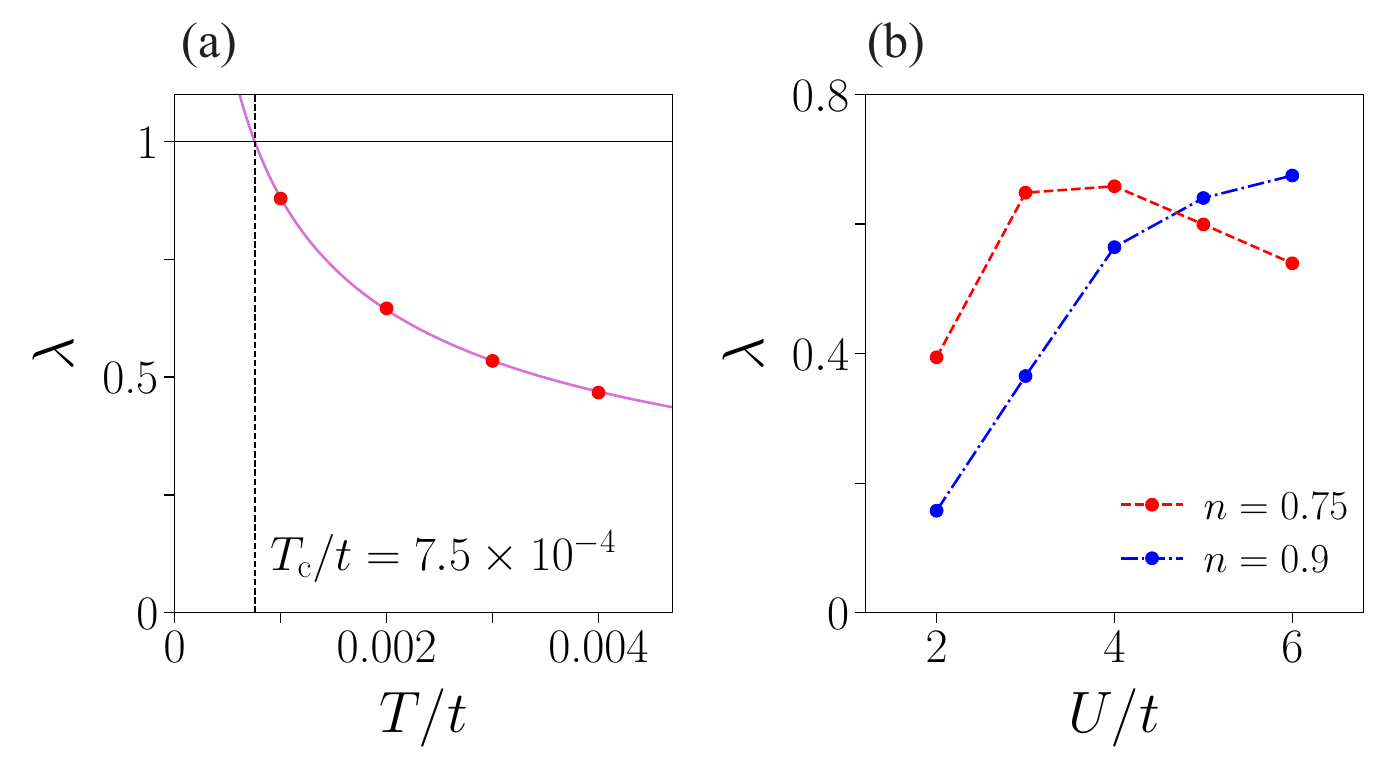}
    \caption{(a)~Temperature $T$ dependence of $\lambda$ for $n=0.75$, $\alpha=1$, and $U/t=4$. 
    $8192\times2$ Matsubara frequencies were used. 
    The superconducting transition temperature $T_{\mathrm{c}}$ is estimated by extrapolation.
    (b)~On-site Coulomb interaction $U$ dependence of $\lambda$ for $T/t=0.002$ and $\alpha=1$.}
    \label{fig:lambda_details}
\end{center}
\end{figure}

We present additional results obtained from the FLEX calculation and the linearized Eliashberg equation.
In Fig.~\ref{fig:lambda_details}, we show the temperature $T$ and on-site Coulomb interaction $U$ dependence of the eigenvalue $\lambda$ of the linearized Eliashberg equation.
In Fig.~\ref{fig:lambda_details}(a), we estimate the superconducting transition temperature $T_{\mathrm{c}}$ based on $\lambda$ calculated at $T/t = 0.004$, $0.003$, $0.002$, and $0.001$ for $\alpha = 1$ and $n = 0.75$, where the largest $\lambda$ is obtained in the main text.
Here, we used $8192\times2$ Matsubara frequencies, which are twice the number used in the main text.
At $T/t = 0.002$, the eigenvalue $\lambda$ obtained with $4096\times2$ Matsubara frequencies shows quantitative agreement, indicating that the number of frequencies used in the main text is sufficient for convergence at this temperature.
As shown in Fig.~\ref{fig:lambda_details}(a), $\lambda$ increases monotonically with decreasing temperature and reaches $\lambda = 1$ at $T = 7.5\times10^{-4}\,t$, which is identified as the superconducting transition temperature $T_{\mathrm{c}}$.
In Fig.~\ref{fig:lambda_details}(b), we show the $U$ dependence of $\lambda$ for band fillings $n = 0.75$ and $n = 0.9$ at $\alpha = 1$.
For $n = 0.9$, $\lambda$ increases monotonically with increasing $U$ in the range $U/t \leq 6$, whereas for $n = 0.75$, $\lambda$ takes its maximum around $U/t = 3$--$4$.

\begin{figure}[t]
\begin{center}
    \includegraphics[width=\columnwidth]{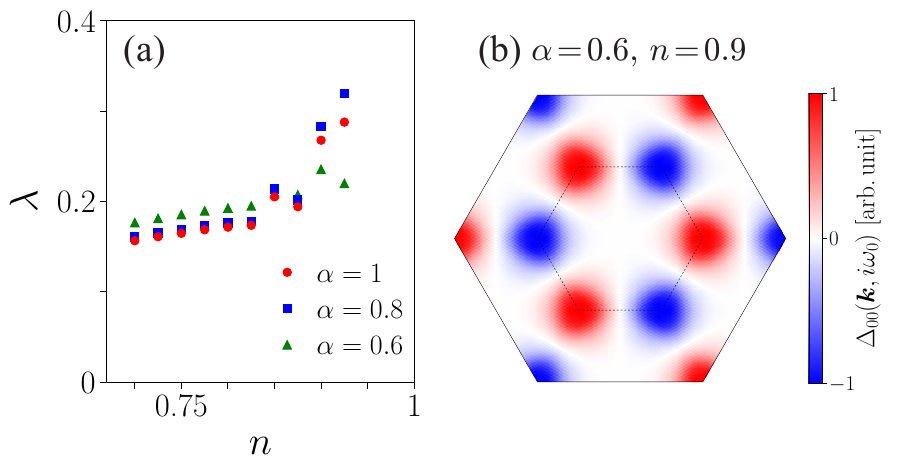}
    \caption{
    (a)~Band-filling $n$ dependence of the eigenvalue $\lambda$ of the linearized Eliashberg equation for the spin-triplet pairing interaction.
    (b)~Gap function $\Delta(\bm{k}, i\omega_0)$ at the lowest Matsubara frequency $i\omega_0$ for $\alpha = 0.6$ and $n = 0.9$, where $\lambda_{\rm tirplet}>\lambda_{\rm singlet}$ holds at $T/t=0.002$.
    The $f$-wave symmetry is obtained for all other parameter sets as well.
    } 
    \label{fig:triplet}
\end{center}
\end{figure}

Next, we examine the possibility of spin-triplet superconductivity.
Several studies on the repulsive Hubbard model on a honeycomb lattice have suggested the presence of fluctuations favoring $f$-wave superconductivity near half-filling~\cite{Honerkamp_2008,Raghu_2010,Nandkishore_2014}.
To investigate the spin-triplet superconductivity mediated by spin fluctuations, we replace the pairing interaction in the linearized Eliashberg equation~(\ref{eq:Eliash_eq}) with that for the spin-triplet channel,
\begin{align}\label{eq:pairing_interaction_triplet}
    \Gamma(\bm{q},i\Omega_m)
    &=-\frac12U^{\rm s}\chi^{\rm s}(\bm{q},i\Omega_m)U^{\rm s} 
    - \frac12U^{\rm c}\chi^{\rm c}(\bm{q},i\Omega_m)U^{\rm c} \notag\\
    &\quad- \frac12\left(U^{\rm s}\!-\!U^{\rm c}\right).
\end{align}
In Fig.~\ref{fig:triplet}(a), we show the band filling $n$ dependence of the eigenvalue $\lambda$ of the linearized Eliashberg equation with a pairing interaction given in Eq.~(\ref{eq:pairing_interaction_triplet}).
In all cases, we obtain an $f$-wave symmetry as shown in Fig.~\ref{fig:triplet}(b).
The eigenvalue $\lambda$ tends to increase toward $1/3$-filling.
However, the $d$-wave state remains dominant for $\alpha = 1$ and $\alpha = 0.8$.
For $\alpha = 0.6$, where spin-singlet $\lambda$ is strongly suppressed near $1/3$-filling, an $f$-wave symmetry may be favored.
It should be noted, however, that magnetic order is likely to be favored over superconductivity because $\lambda$ is small and the Stoner factor is close to $1$.

\section{All matrix elements of the dynamical spin susceptibility\label{app:chi_S_dynamical}}

\begin{figure*}[t]
\begin{center}
    \includegraphics[width=0.9\linewidth]{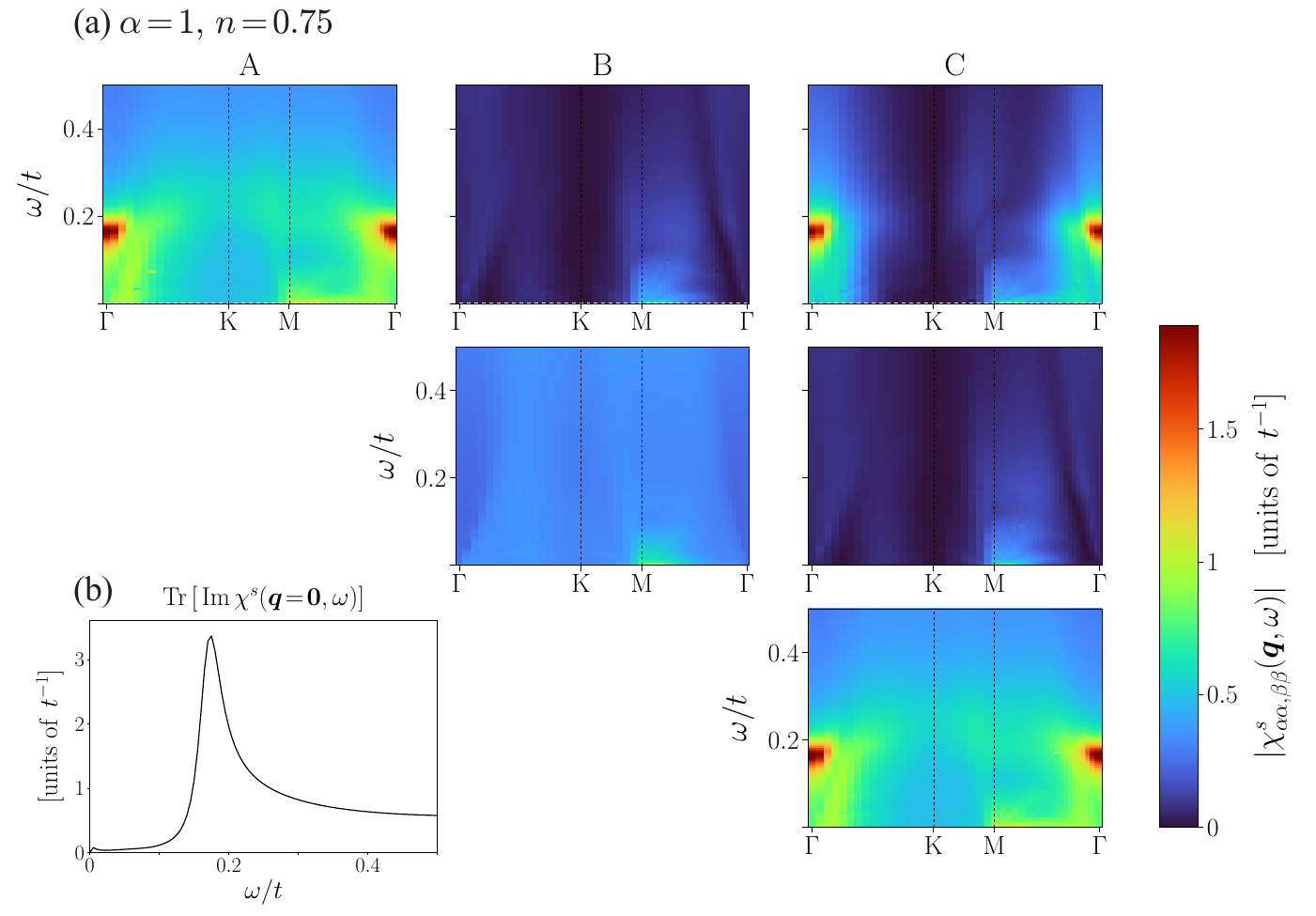}
    \caption{
    (a)~$\alpha\beta$ components ($\alpha,\beta\in\{{\rm A,B,C}\}$) of the dynamical spin susceptibility $\chi^{\rm s}_{\alpha\alpha,\beta\beta}(\bm{q},\omega)$ calculated at $\alpha=1$ and $n=0.75$.
    The AC component (top right panel) is the same as that shown in Fig.~\ref{fig:spec_suscep_schematic}(a).
    (b)~Trace of the imaginary part of the dynamical spin susceptibility at $\bm{q}=0$.
    } 
    \label{fig:suscep_matrix}
\end{center}
\end{figure*}

In Fig.~\ref{fig:suscep_matrix}(a), we show all matrix elements of the dynamical spin susceptibility $\chi^{\rm s}(\bm{q},\omega)$ in the sublattice representation for $\alpha=1$ and $n=0.75$.
The static ($\omega=0$) components exhibit large spectral weight along the $\Gamma$--M line.
In contrast, the dynamical components display distinct behaviors depending on the matrix elements.
The diagonal AA and CC components, as well as the off-diagonal AC component, have the largest weight around the $\Gamma$ point ($\bm{q}=\bm{0}$) in the finite-energy region.
On the other hand, the components involving the hub site (BB, AB, and BC) exhibit no notable features in the finite-energy region.

The quantity measured in experiments such as inelastic neutron scattering often corresponds to the imaginary part of the dynamical spin susceptibility.
Therefore, in Fig.~\ref{fig:suscep_matrix}(b), we present the sum of the imaginary parts of the diagonal elements of $\chi^{\rm s}(\bm{q},\omega)$ at $\bm{q}=\bm{0}$.
At $n=0.75$, low-energy spin fluctuations are strongly suppressed, and a pronounced peak emerges near the energy corresponding to the flat-band position.
As the Fermi level approaches the flat band, the peak shifts toward lower energies.

\bibliography{reference}

@Article{Aoki_2020_review,
author={Aoki, Hideo},
title={{Theoretical Possibilities for Flat Band Superconductivity}},
journal={J. Supercond. Novel Magn.},
year={2020},
month={Aug},
day={01},
volume={33},
number={8},
pages={2341-2346},
issn={1557-1947},
doi={10.1007/s10948-020-05474-6},
url={https://doi.org/10.1007/s10948-020-05474-6}
}

@article{Aoki_2025_review,
author = {Hideo Aoki},
title = {{Flat bands in condensed-matter systems -- perspective for magnetism and superconductivity}},
journal = {Contemp. Phys.},
volume = {66},
number = {1-4},
pages = {1--38},
year = {2025},
publisher = {Taylor \& Francis},
doi = {10.1080/00107514.2025.2550105},
URL = {https://doi.org/10.1080/00107514.2025.2550105},
}

@article{Bang_2014,
doi = {10.1088/1367-2630/16/2/023029},
url = {https://doi.org/10.1088/1367-2630/16/2/023029},
year = {2014},
month = {feb},
publisher = {IOP Publishing},
volume = {16},
number = {2},
pages = {023029},
author = {Bang, Yunkyu},
title = {{A shadow gap in the over-doped (Ba$_{1-x}$K$_x$)Fe$_2$As$_2$ compound}},
journal = {New J. Phys.},
}

@article{Black-Schaffer_2014_review,
doi = {10.1088/0953-8984/26/42/423201},
url = {https://doi.org/10.1088/0953-8984/26/42/423201},
year = {2014},
month = {sep},
publisher = {IOP Publishing},
volume = {26},
number = {42},
pages = {423201},
author = {Black-Schaffer, Annica M and Honerkamp, Carsten},
title = {{Chiral d-wave superconductivity in doped graphene}},
journal = {J. Phys.: Condens. Matter},
}

@article{Ghosh_2021_review,
doi = {10.1088/1361-648X/abaa06},
url = {https://doi.org/10.1088/1361-648X/abaa06},
year = {2020},
month = {oct},
publisher = {IOP Publishing},
volume = {33},
number = {3},
pages = {033001},
author = {Ghosh, Sudeep Kumar and Smidman, Michael and Shang, Tian and Annett, James F and Hillier, Adrian D and Quintanilla, Jorge and Yuan, Huiqiu},
title = {{Recent progress on superconductors with time-reversal symmetry breaking}},
journal = {J. Phys.: Condens. Matter},
}

@article{Kallin_2016_review,
doi = {10.1088/0034-4885/79/5/054502},
url = {https://doi.org/10.1088/0034-4885/79/5/054502},
year = {2016},
month = {apr},
publisher = {IOP Publishing},
volume = {79},
number = {5},
pages = {054502},
author = {Kallin, Catherine and Berlinsky, John},
title = {{Chiral superconductors}},
journal = {Rep. Prog. Phys.},
}

@article{Leykam_2018_review,
author = {Daniel Leykam and Alexei Andreanov and Sergej Flach},
title = {{Artificial flat band systems: from lattice models to experiments}},
journal = {Adv. Phys.: X},
volume = {3},
number = {1},
pages = {1473052},
year = {2018},
publisher = {Taylor \& Francis},
doi = {10.1080/23746149.2018.1473052},
URL = {https://doi.org/10.1080/23746149.2018.1473052},
}

@article{Micnas_1990_review,
  title = {{Superconductivity in narrow-band systems with local nonretarded attractive interactions}},
  author = {Micnas, R. and Ranninger, J. and Robaszkiewicz, S.},
  journal = {Rev. Mod. Phys.},
  volume = {62},
  issue = {1},
  pages = {113--171},
  numpages = {0},
  year = {1990},
  month = {Jan},
  publisher = {American Physical Society},
  doi = {10.1103/RevModPhys.62.113},
  url = {https://link.aps.org/doi/10.1103/RevModPhys.62.113}
}

@article{Nayak_2008_review,
  title = {{Non-Abelian anyons and topological quantum computation}},
  author = {Nayak, Chetan and Simon, Steven H. and Stern, Ady and Freedman, Michael and Das Sarma, Sankar},
  journal = {Rev. Mod. Phys.},
  volume = {80},
  issue = {3},
  pages = {1083--1159},
  numpages = {0},
  year = {2008},
  month = {Sep},
  publisher = {American Physical Society},
  doi = {10.1103/RevModPhys.80.1083},
  url = {https://link.aps.org/doi/10.1103/RevModPhys.80.1083}
}

@article{Sato_2017_review,
doi = {10.1088/1361-6633/aa6ac7},
url = {https://doi.org/10.1088/1361-6633/aa6ac7},
year = {2017},
month = {may},
publisher = {IOP Publishing},
volume = {80},
number = {7},
pages = {076501},
author = {Sato, Masatoshi and Ando, Yoichi},
title = {{Topological superconductors: a review}},
journal = {Rep. Prog. Phys.},
}

@article{Aida_2024,
  title = {{Theoretical study of spin-fluctuation-mediated superconductivity in two-dimensional Hubbard models with an incipient flat band}},
  author = {Aida, Tetsuaki and Matsumoto, Karin and Ogura, Daisuke and Ochi, Masayuki and Kuroki, Kazuhiko},
  journal = {Phys. Rev. B},
  volume = {110},
  issue = {5},
  pages = {054516},
  numpages = {8},
  year = {2024},
  month = {Aug},
  publisher = {American Physical Society},
  doi = {10.1103/PhysRevB.110.054516},
  url = {https://link.aps.org/doi/10.1103/PhysRevB.110.054516}
}

@article{Anderson_1950,
  title = {{Antiferromagnetism. Theory of Superexchange Interaction}},
  author = {Anderson, P. W.},
  journal = {Phys. Rev.},
  volume = {79},
  issue = {2},
  pages = {350--356},
  numpages = {0},
  year = {1950},
  month = {Jul},
  publisher = {American Physical Society},
  doi = {10.1103/PhysRev.79.350},
  url = {https://link.aps.org/doi/10.1103/PhysRev.79.350}
}

@article{Anderson_1959,
  title = {{New Approach to the Theory of Superexchange Interactions}},
  author = {Anderson, P. W.},
  journal = {\emph{ibid.}},
  volume = {115},
  issue = {1},
  pages = {2--13},
  numpages = {0},
  year = {1959},
  month = {Jul},
  publisher = {American Physical Society},
  doi = {10.1103/PhysRev.115.2},
  url = {https://link.aps.org/doi/10.1103/PhysRev.115.2}
}

@article{Aoki_1996,
  title = {Hofstadter butterflies for flat bands},
  author = {Aoki, Hideo and Ando, Masato and Matsumura, Hajime},
  journal = {Phys. Rev. B},
  volume = {54},
  issue = {24},
  pages = {R17296--R17299},
  numpages = {0},
  year = {1996},
  month = {Dec},
  publisher = {American Physical Society},
  doi = {10.1103/PhysRevB.54.R17296},
  url = {https://link.aps.org/doi/10.1103/PhysRevB.54.R17296}
}

@article{Bang_2016,
doi = {10.1088/1367-2630/18/11/113054},
url = {https://doi.org/10.1088/1367-2630/18/11/113054},
year = {2016},
month = {nov},
publisher = {IOP Publishing},
volume = {18},
number = {11},
pages = {113054},
author = {Bang, Yunkyu},
title = {{Pairing mechanism of heavily electron doped FeSe systems: dynamical tuning of the pairing cutoff energy}},
journal = {New J. Phys.},
}

@article{Black-Schaffer_2007,
  title = {{Resonating valence bonds and mean-field $d$-wave superconductivity in graphite}},
  author = {Black-Schaffer, Annica M. and Doniach, Sebastian},
  journal = {Phys. Rev. B},
  volume = {75},
  issue = {13},
  pages = {134512},
  numpages = {10},
  year = {2007},
  month = {Apr},
  publisher = {American Physical Society},
  doi = {10.1103/PhysRevB.75.134512},
  url = {https://link.aps.org/doi/10.1103/PhysRevB.75.134512}
}

@article{Bickers_1989,
  title = {{Conserving Approximations for Strongly Correlated Electron Systems: Bethe-Salpeter Equation and Dynamics for the Two-Dimensional Hubbard Model}},
  author = {Bickers, N. E. and Scalapino, D. J. and White, S. R.},
  journal = {Phys. Rev. Lett.},
  volume = {62},
  issue = {8},
  pages = {961--964},
  numpages = {0},
  year = {1989},
  month = {Feb},
  publisher = {American Physical Society},
  doi = {10.1103/PhysRevLett.62.961},
  url = {https://link.aps.org/doi/10.1103/PhysRevLett.62.961}
}

@Article{Charnukha_2015,
author={Charnukha, A. and Evtushinsky, D. V. and Matt, C. E. and Xu, N. and Shi, M. and B{\"u}chner, B. and Zhigadlo, N. D. and Batlogg, B. and Borisenko, S. V.},
title={{High-temperature superconductivity from fine-tuning of Fermi-surface singularities in iron oxypnictides}},
journal={Sci. Rep.},
year={2015},
month={Dec},
day={18},
volume={5},
number={1},
pages={18273},
issn={2045-2322},
doi={10.1038/srep18273},
url={https://doi.org/10.1038/srep18273}
}

@Article{Y-Cao_2018,
author={Cao, Yuan and Fatemi, Valla and Fang, Shiang and Watanabe, Kenji and Taniguchi, Takashi and Kaxiras, Efthimios and Jarillo-Herrero, Pablo},
title={{Unconventional superconductivity in magic-angle graphene superlattices}},
journal={Nature},
year={2018},
month={Apr},
day={01},
volume={556},
number={7699},
pages={43-50},
issn={1476-4687},
doi={10.1038/nature26160},
url={https://doi.org/10.1038/nature26160}
}

@article{X-Chen_2015,
  title = {{Electron pairing in the presence of incipient bands in iron-based superconductors}},
  author = {Chen, Xiao and Maiti, S. and Linscheid, A. and Hirschfeld, P. J.},
  journal = {Phys. Rev. B},
  volume = {92},
  issue = {22},
  pages = {224514},
  numpages = {14},
  year = {2015},
  month = {Dec},
  publisher = {American Physical Society},
  doi = {10.1103/PhysRevB.92.224514},
  url = {https://link.aps.org/doi/10.1103/PhysRevB.92.224514}
}

@article{Crepieux_2023,
  title = {{Superconductivity in monolayer and few-layer graphene. II. Topological edge states and Chern numbers}},
  author = {Cr\'epieux, Adeline and Pangburn, Emile and Haurie, Louis and Awoga, Oladunjoye A. and Black-Schaffer, Annica M. and Sedlmayr, Nicholas and P\'epin, Catherine and Bena, Cristina},
  journal = {Phys. Rev. B},
  volume = {108},
  issue = {13},
  pages = {134515},
  numpages = {14},
  year = {2023},
  month = {Oct},
  publisher = {American Physical Society},
  doi = {10.1103/PhysRevB.108.134515},
  url = {https://link.aps.org/doi/10.1103/PhysRevB.108.134515}
}

@article{Dahm_1995,
  title = {{Quasiparticle and Spin Excitation Spectra in the Normal and $d$-Wave Superconducting State of the Two-Dimensional Hubbard Model}},
  author = {Dahm, T. and Tewordt, L.},
  journal = {Phys. Rev. Lett.},
  volume = {74},
  issue = {5},
  pages = {793--796},
  numpages = {0},
  year = {1995},
  month = {Jan},
  publisher = {American Physical Society},
  doi = {10.1103/PhysRevLett.74.793},
  url = {https://link.aps.org/doi/10.1103/PhysRevLett.74.793}
}

@article{Emery_1987,
  title = {{Theory of high-${\mathrm{T}}_{\mathrm{c}}$ superconductivity in oxides}},
  author = {Emery, V. J.},
  journal = {Phys. Rev. Lett.},
  volume = {58},
  issue = {26},
  pages = {2794--2797},
  numpages = {0},
  year = {1987},
  month = {Jun},
  publisher = {American Physical Society},
  doi = {10.1103/PhysRevLett.58.2794},
  url = {https://link.aps.org/doi/10.1103/PhysRevLett.58.2794}
}

@article{Fukui_2005,
author = {Fukui ,Takahiro and Hatsugai ,Yasuhiro and Suzuki ,Hiroshi},
title = {{Chern Numbers in Discretized Brillouin Zone: Efficient Method of Computing (Spin) Hall Conductances}},
journal = {J. Phys. Soc. Jpn.},
volume = {74},
number = {6},
pages = {1674-1677},
year = {2005},
doi = {10.1143/JPSJ.74.1674},
URL = { https://doi.org/10.1143/JPSJ.74.1674},
}

@article{Gonzalez_2008,
  title = {{Kohn-Luttinger superconductivity in graphene}},
  author = {Gonz\'alez, J.},
  journal = {Phys. Rev. B},
  volume = {78},
  issue = {20},
  pages = {205431},
  numpages = {6},
  year = {2008},
  month = {Nov},
  publisher = {American Physical Society},
  doi = {10.1103/PhysRevB.78.205431},
  url = {https://link.aps.org/doi/10.1103/PhysRevB.78.205431}
}

@Article{Han_2025,
author={Han, Tonghang and Lu, Zhengguang and Hadjri, Zach and Shi, Lihan and Wu, Zhenghan and Xu, Wei and Yao, Yuxuan and Cotten, Armel A. and Sharifi Sedeh, Omid and Weldeyesus, Henok and Yang, Jixiang and Seo, Junseok and Ye, Shenyong and Zhou, Muyang and Liu, Haoyang and Shi, Gang and Hua, Zhenqi and Watanabe, Kenji and Taniguchi, Takashi and Xiong, Peng and Zumb{\"u}hl, Dominik M. and Fu, Liang and Ju, Long},
title={{Signatures of chiral superconductivity in rhombohedral graphene}},
journal={Nature (London)},
year={2025},
month={Jul},
day={01},
volume={643},
number={8072},
pages={654-661},
issn={1476-4687},
doi={10.1038/s41586-025-09169-7},
url={https://doi.org/10.1038/s41586-025-09169-7}
}

@article{Holmvall_2023a,
  title = {{Coreless vortices as direct signature of chiral $d$-wave superconductivity}},
  author = {Holmvall, P. and Black-Schaffer, A. M.},
  journal = {Phys. Rev. B},
  volume = {108},
  issue = {10},
  pages = {L100506},
  numpages = {9},
  year = {2023},
  month = {Sep},
  publisher = {American Physical Society},
  doi = {10.1103/PhysRevB.108.L100506},
  url = {https://link.aps.org/doi/10.1103/PhysRevB.108.L100506}
}

@article{Holmvall_2023b,
  title = {{Robust and tunable coreless vortices and fractional vortices in chiral $d$-wave superconductors}},
  author = {Holmvall, P. and Wall-Wennerdal, N. and Black-Schaffer, A. M.},
  journal = {Phys. Rev. B},
  volume = {108},
  issue = {9},
  pages = {094511},
  numpages = {24},
  year = {2023},
  month = {Sep},
  publisher = {American Physical Society},
  doi = {10.1103/PhysRevB.108.094511},
  url = {https://link.aps.org/doi/10.1103/PhysRevB.108.094511}
}

@article{Holmvall_2023c,
  title = {{Enhanced chiral edge currents and orbital magnetic moment in chiral $d$-wave superconductors from mesoscopic finite-size effects}},
  author = {Holmvall, P. and Black-Schaffer, A. M.},
  journal = {Phys. Rev. B},
  volume = {108},
  issue = {17},
  pages = {174505},
  numpages = {24},
  year = {2023},
  month = {Nov},
  publisher = {American Physical Society},
  doi = {10.1103/PhysRevB.108.174505},
  url = {https://link.aps.org/doi/10.1103/PhysRevB.108.174505}
}

@article{Holmvall_2025,
  title = {{Designing edge currents using mesoscopic patterning in chiral $d$-wave superconductors}},
  author = {Holmvall, Patric and Black-Schaffer, Annica M.},
  journal = {Phys. Rev. B},
  volume = {111},
  issue = {18},
  pages = {184505},
  numpages = {19},
  year = {2025},
  month = {May},
  publisher = {American Physical Society},
  doi = {10.1103/PhysRevB.111.184505},
  url = {https://link.aps.org/doi/10.1103/PhysRevB.111.184505}
}

@article{Horiguchi_1974,
    author = {Horiguchi, T. and Chen, C. C.},
    title = {{Lattice Green's function for the diced lattice}},
    journal = {J. Math. Phys.},
    volume = {15},
    number = {5},
    pages = {659-660},
    year = {1974},
    month = {05},
    issn = {0022-2488},
    doi = {10.1063/1.1666703},
    url = {https://doi.org/10.1063/1.1666703},
}

@article{Kato_2020,
  title = {{Many-variable variational Monte Carlo study of superconductivity in two-band Hubbard models with an incipient band}},
  author = {Kato, Daichi and Kuroki, Kazuhiko},
  journal = {Phys. Rev. Res.},
  volume = {2},
  issue = {2},
  pages = {023156},
  numpages = {9},
  year = {2020},
  month = {May},
  publisher = {American Physical Society},
  doi = {10.1103/PhysRevResearch.2.023156},
  url = {https://link.aps.org/doi/10.1103/PhysRevResearch.2.023156}
}

@article{Kitamura_2024,
  title = {{Spin-Triplet Superconductivity from Quantum-Geometry-Induced Ferromagnetic Fluctuation}},
  author = {Kitamura, Taisei and Daido, Akito and Yanase, Youichi},
  journal = {Phys. Rev. Lett.},
  volume = {132},
  issue = {3},
  pages = {036001},
  numpages = {6},
  year = {2024},
  month = {Jan},
  publisher = {American Physical Society},
  doi = {10.1103/PhysRevLett.132.036001},
  url = {https://link.aps.org/doi/10.1103/PhysRevLett.132.036001}
}

@article{Kobayashi_2016,
  title = {{Superconductivity in repulsively interacting fermions on a diamond chain: Flat-band-induced pairing}},
  author = {Kobayashi, Keita and Okumura, Masahiko and Yamada, Susumu and Machida, Masahiko and Aoki, Hideo},
  journal = {Phys. Rev. B},
  volume = {94},
  issue = {21},
  pages = {214501},
  numpages = {7},
  year = {2016},
  month = {Dec},
  publisher = {American Physical Society},
  doi = {10.1103/PhysRevB.94.214501},
  url = {https://link.aps.org/doi/10.1103/PhysRevB.94.214501}
}

@article{Kouchi_2022,
  title = {{Enhanced superconductivity and moderate spin fluctuations suppressed at low energies in heavily electron-doped La1111-based superconductor}},
  author = {Kouchi, T. and Nishioka, S. and Suzuki, K. and Yashima, M. and Mukuda, H. and Kawashima, T. and Tsuji, H. and Kuroki, K. and Miyasaka, S. and Tajima, S.},
  journal = {Phys. Rev. B},
  volume = {105},
  issue = {14},
  pages = {144510},
  numpages = {6},
  year = {2022},
  month = {Apr},
  publisher = {American Physical Society},
  doi = {10.1103/PhysRevB.105.144510},
  url = {https://link.aps.org/doi/10.1103/PhysRevB.105.144510}
}

@article{Kuroki_2001,
  title = {{Spin-triplet superconductivity in repulsive Hubbard models with disconnected Fermi surfaces: A case study on triangular and honeycomb lattices}},
  author = {Kuroki, Kazuhiko and Arita, Ryotaro},
  journal = {Phys. Rev. B},
  volume = {63},
  issue = {17},
  pages = {174507},
  numpages = {5},
  year = {2001},
  month = {Apr},
  publisher = {American Physical Society},
  doi = {10.1103/PhysRevB.63.174507},
  url = {https://link.aps.org/doi/10.1103/PhysRevB.63.174507}
}

@article{Kuroki_2005,
  title = {{High-${T}_{c}$ superconductivity due to coexisting wide and narrow bands: A fluctuation exchange study of the Hubbard ladder as a test case}},
  author = {Kuroki, Kazuhiko and Higashida, Takafumi and Arita, Ryotaro},
  journal = {Phys. Rev. B},
  volume = {72},
  issue = {21},
  pages = {212509},
  numpages = {4},
  year = {2005},
  month = {Dec},
  publisher = {American Physical Society},
  doi = {10.1103/PhysRevB.72.212509},
  url = {https://link.aps.org/doi/10.1103/PhysRevB.72.212509}
}

@article{Kuroki_2010,
  title = {{Spin-fluctuation-mediated $d+i{d}^{\ensuremath{'}}$ pairing mechanism in doped $\ensuremath{\beta}\text{\ensuremath{-}}M\text{NCl}$ $(M=\text{Hf},\text{Zr})$ superconductors}},
  author = {Kuroki, Kazuhiko},
  journal = {Phys. Rev. B},
  volume = {81},
  issue = {10},
  pages = {104502},
  numpages = {7},
  year = {2010},
  month = {Mar},
  publisher = {American Physical Society},
  doi = {10.1103/PhysRevB.81.104502},
  url = {https://link.aps.org/doi/10.1103/PhysRevB.81.104502}
}

@article{C-Lu_2018, 
  title = {{$d+id$ chiral superconductivity in a triangular lattice from trigonal bipyramidal complexes}}, 
  author = {Lu, Chen and Zhang, Li-Da and Wu, Xianxin and Yang, Fan and Hu, Jiangping},
  journal = {Phys. Rev. B},
  volume = {97},
  issue = {16},
  pages = {165110},
  numpages = {7},
  year = {2018},
  month = {Apr},
  publisher = {American Physical Society},
  doi = {10.1103/PhysRevB.97.165110},
  url = {https://link.aps.org/doi/10.1103/PhysRevB.97.165110}
}

@article{T-Ma_2011,
  title = {{Pairing in graphene: A quantum Monte Carlo study}},
  author = {Ma, Tianxing and Huang, Zhongbing and Hu, Feiming and Lin, Hai-Qing},
  journal = {Phys. Rev. B},
  volume = {84},
  issue = {12},
  pages = {121410},
  numpages = {4},
  year = {2011},
  month = {Sep},
  publisher = {American Physical Society},
  doi = {10.1103/PhysRevB.84.121410},
  url = {https://link.aps.org/doi/10.1103/PhysRevB.84.121410}
}

@article{Lieb_1989,
  title = {{Two theorems on the Hubbard model}},
  author = {Lieb, Elliott H.},
  journal = {Phys. Rev. Lett.},
  volume = {62},
  issue = {10},
  pages = {1201--1204},
  numpages = {0},
  year = {1989},
  month = {Mar},
  publisher = {American Physical Society},
  doi = {10.1103/PhysRevLett.62.1201},
  url = {https://link.aps.org/doi/10.1103/PhysRevLett.62.1201}
}

@article{Linscheid_2016,
  title = {{High ${T}_{c}$ via Spin Fluctuations from Incipient Bands: Application to Monolayers and Intercalates of FeSe}},
  author = {Linscheid, A. and Maiti, S. and Wang, Y. and Johnston, S. and Hirschfeld, P. J.},
  journal = {Phys. Rev. Lett.},
  volume = {117},
  issue = {7},
  pages = {077003},
  numpages = {5},
  year = {2016},
  month = {Aug},
  publisher = {American Physical Society},
  doi = {10.1103/PhysRevLett.117.077003},
  url = {https://link.aps.org/doi/10.1103/PhysRevLett.117.077003}
}

@article{Matsumoto_2018,
  title = {{Wide applicability of high-${T}_{c}$ pairing originating from coexisting wide and incipient narrow bands in quasi-one-dimensional systems}},
  author = {Matsumoto, Karin and Ogura, Daisuke and Kuroki, Kazuhiko},
  journal = {Phys. Rev. B},
  volume = {97},
  issue = {1},
  pages = {014516},
  numpages = {11},
  year = {2018},
  month = {Jan},
  publisher = {American Physical Society},
  doi = {10.1103/PhysRevB.97.014516},
  url = {https://link.aps.org/doi/10.1103/PhysRevB.97.014516}
}

@article{Matsumoto_2020,
author = {Matsumoto ,Karin and Ogura ,Daisuke and Kuroki ,Kazuhiko},
title = {{Strongly Enhanced Superconductivity Due to Finite Energy Spin Fluctuations Induced by an Incipient Band: A FLEX Study on the Bilayer Hubbard Model with Vertical and Diagonal Interlayer Hoppings}},
journal = {J. Phys. Soc. Jpn.},
volume = {89},
number = {4},
pages = {044709},
year = {2020},
doi = {10.7566/JPSJ.89.044709},
URL = {https://doi.org/10.7566/JPSJ.89.044709},
}

@article{Matthew_2024,
  title = {{Chern number landscape of spin-orbit coupled chiral superconductors}},
  author = {Bunney, Matthew and Beyer, Jacob and Thomale, Ronny and Honerkamp, Carsten and Rachel, Stephan},
  journal = {Phys. Rev. B},
  volume = {110},
  issue = {16},
  pages = {L161103},
  numpages = {6},
  year = {2024},
  month = {Oct},
  publisher = {American Physical Society},
  doi = {10.1103/PhysRevB.110.L161103},
  url = {https://link.aps.org/doi/10.1103/PhysRevB.110.L161103}
}

@article{Maier_2019,
  title = {{Effective pairing interaction in a system with an incipient band}},
  author = {Maier, T. A. and Mishra, V. and Balduzzi, G. and Scalapino, D. J.},
  journal = {Phys. Rev. B},
  volume = {99},
  issue = {14},
  pages = {140504},
  numpages = {5},
  year = {2019},
  month = {Apr},
  publisher = {American Physical Society},
  doi = {10.1103/PhysRevB.99.140504},
  url = {https://link.aps.org/doi/10.1103/PhysRevB.99.140504}
}

@Article{Miao_2015,
author={Miao, H. and Qian, T. and Shi, X. and Richard, P. and Kim, T. K. and Hoesch, M. and Xing, L. Y. and Wang, X.-C. and Jin, C.-Q. and Hu, J.-P. and Ding, H.},
title={{Observation of strong electron pairing on bands without Fermi surfaces in LiFe$_{1-x}$Co$_x$As}},
journal={Nat. Commun.},
year={2015},
month={Jan},
day={13},
volume={6},
number={1},
pages={6056},
issn={2041-1723},
doi={10.1038/ncomms7056},
url={https://doi.org/10.1038/ncomms7056}
}

@article{Mielke_1991a,
doi = {10.1088/0305-4470/24/2/005},
url = {https://doi.org/10.1088/0305-4470/24/2/005},
year = {1991},
month = {jan},
publisher = {},
volume = {24},
number = {2},
pages = {L73},
author = {A Mielke},
title = {{Ferromagnetic ground states for the Hubbard model on line graphs}},
journal = {J. Phys. A: Math. Gen.},
}

@article{Mielke_1991b,
doi = {10.1088/0305-4470/24/14/018},
url = {https://doi.org/10.1088/0305-4470/24/14/018},
year = {1991},
month = {jul},
publisher = {},
volume = {24},
number = {14},
pages = {3311},
author = {A Mielke},
title = {{Ferromagnetism in the Hubbard model on line graphs and further considerations}},
journal = {\emph{ibid.}},
}

@Article{Mishra_2016,
    author={Mishra, Vivek and Scalapino, Douglas J. and Maier, Thomas A.},
    title={{s{\textpm} pairing near a Lifshitz transition}},
    journal={Sci. Rep.},
    year={2016},
    month={Aug},
    day={26},
    volume={6},
    number={1},
    pages={32078},
    issn={2045-2322},
    doi={10.1038/srep32078},
    url={https://doi.org/10.1038/srep32078}
}

@article{T-Nakano_2007,
  title = {{Superconductivity due to spin fluctuations originating from multiple Fermi surfaces in the double chain superconductor ${\mathrm{Pr}}_{2}{\mathrm{Ba}}_{4}{\mathrm{Cu}}_{7}{\mathrm{O}}_{15\ensuremath{-}\ensuremath{\delta}}$}},
  author = {Nakano, Tsuguhito and Kuroki, Kazuhiko and Onari, Seiichiro},
  journal = {Phys. Rev. B},
  volume = {76},
  issue = {1},
  pages = {014515},
  numpages = {7},
  year = {2007},
  month = {Jul},
  publisher = {American Physical Society},
  doi = {10.1103/PhysRevB.76.014515},
  url = {https://link.aps.org/doi/10.1103/PhysRevB.76.014515}
}

@article{H-Nakano_2017,
author = {Nakano ,Hiroki and Sakai ,T\^{o}ru},
title = {{Ferrimagnetism in the Spin-1/2 Heisenberg Antiferromagnet on a Distorted Triangular Lattice}},
journal = {J. Phys. Soc. Jpn.},
volume = {86},
number = {6},
pages = {063702},
year = {2017},
doi = {10.7566/JPSJ.86.063702},
URL = {https://doi.org/10.7566/JPSJ.86.063702},
}

@article{Nakata_2017,
  title = {{Finite-energy spin fluctuations as a pairing glue in systems with coexisting electron and hole bands}},
  author = {Nakata, Masahiro and Ogura, Daisuke and Usui, Hidetomo and Kuroki, Kazuhiko},
  journal = {Phys. Rev. B},
  volume = {95},
  issue = {21},
  pages = {214509},
  numpages = {6},
  year = {2017},
  month = {Jun},
  publisher = {American Physical Society},
  doi = {10.1103/PhysRevB.95.214509},
  url = {https://link.aps.org/doi/10.1103/PhysRevB.95.214509}
}

@Article{Nandkishore_2012,
author={Nandkishore, Rahul
and Levitov, L. S.
and Chubukov, A. V.},
title={{Chiral superconductivity from repulsive interactions in doped graphene}},
journal={Nat. Phys.},
year={2012},
month={Feb},
day={01},
volume={8},
number={2},
pages={158-163},
issn={1745-2481},
doi={10.1038/nphys2208},
url={https://doi.org/10.1038/nphys2208}
}

@article{Nandkishore_2014,
  title = {{Superconductivity from weak repulsion in hexagonal lattice systems}},
  author = {Nandkishore, Rahul and Thomale, Ronny and Chubukov, Andrey V.},
  journal = {Phys. Rev. B},
  volume = {89},
  issue = {14},
  pages = {144501},
  numpages = {23},
  year = {2014},
  month = {Apr},
  publisher = {American Physical Society},
  doi = {10.1103/PhysRevB.89.144501},
  url = {https://link.aps.org/doi/10.1103/PhysRevB.89.144501}
}

@article{Nishioka_2021,
        author = {Nishioka ,Sotaro and Kouchi ,Takayoshi and Suzuki ,Kazuhiro and Yashima ,Mitsuharu and Mukuda ,Hidekazu and Kodani ,Masashi and Mita ,Kaito and Kakuto ,Takeshi and Lee ,Ji-Hyun and Fujii ,Tatsuo and Kambe ,Takashi},
        title = {{Unconventional Superconductivity and Moderate Spin Fluctuations with Gap at Low Energies in Intercalated Iron Selenide Superconductor Li$_x$(NH$_3$)$_y$Fe$_{2-\delta}$Se$_2$ Probed by $^{77}$Se NMR}},
        journal = {J. Phys. Soc. Jpn.},
        volume = {90},
        number = {12},
        pages = {124709},
        year = {2021},
        doi = {10.7566/JPSJ.90.124709},
        URL = {https://doi.org/10.7566/JPSJ.90.124709},
}

@article{Ogura_2018,
  title = {{Possible high-${T}_{c}$ superconductivity due to incipient narrow bands originating from hidden ladders in Ruddlesden-Popper compounds}},
  author = {Ogura, Daisuke and Aoki, Hideo and Kuroki, Kazuhiko},
  journal = {Phys. Rev. B},
  volume = {96},
  issue = {18},
  pages = {184513},
  numpages = {8},
  year = {2017},
  month = {Nov},
  publisher = {American Physical Society},
  doi = {10.1103/PhysRevB.96.184513},
  url = {https://link.aps.org/doi/10.1103/PhysRevB.96.184513}
}

@article{Pathak_2010,
  title = {{Possible high-temperature superconducting state with a $d+id$ pairing symmetry in doped graphene}},
  author = {Pathak, Sandeep and Shenoy, Vijay B. and Baskaran, G.},
  journal = {Phys. Rev. B},
  volume = {81},
  issue = {8},
  pages = {085431},
  numpages = {5},
  year = {2010},
  month = {Feb},
  publisher = {American Physical Society},
  doi = {10.1103/PhysRevB.81.085431},
  url = {https://link.aps.org/doi/10.1103/PhysRevB.81.085431}
}

@Article{Peotta_2015,
author={Peotta, Sebastiano
and T{\"o}rm{\"a}, P{\"a}ivi},
title={Superfluidity in topologically nontrivial flat bands},
journal={Nat. Commun.},
year={2015},
month={Nov},
day={20},
volume={6},
number={1},
pages={8944},
issn={2041-1723},
doi={10.1038/ncomms9944},
url={https://doi.org/10.1038/ncomms9944}
}

@article{Raghu_2010,
  title = {{Superconductivity in the repulsive Hubbard model: An asymptotically exact weak-coupling solution}},
  author = {Raghu, S. and Kivelson, S. A. and Scalapino, D. J.},
  journal = {Phys. Rev. B},
  volume = {81},
  issue = {22},
  pages = {224505},
  numpages = {17},
  year = {2010},
  month = {Jun},
  publisher = {American Physical Society},
  doi = {10.1103/PhysRevB.81.224505},
  url = {https://link.aps.org/doi/10.1103/PhysRevB.81.224505}
}

@article{Sakakibara_2024a,
  title = {{Possible High ${T}_{c}$ Superconductivity in ${\mathrm{La}}_{3}{\mathrm{Ni}}_{2}{\mathrm{O}}_{7}$ under High Pressure through Manifestation of a Nearly Half-Filled Bilayer Hubbard Model}},
  author = {Sakakibara, Hirofumi and Kitamine, Naoya and Ochi, Masayuki and Kuroki, Kazuhiko},
  journal = {Phys. Rev. Lett.},
  volume = {132},
  issue = {10},
  pages = {106002},
  numpages = {6},
  year = {2024},
  month = {Mar},
  publisher = {American Physical Society},
  doi = {10.1103/PhysRevLett.132.106002},
  url = {https://link.aps.org/doi/10.1103/PhysRevLett.132.106002}
}

@article{Sakakibara_2024b,
  title = {{Theoretical analysis on the possibility of superconductivity in the trilayer Ruddlesden-Popper nickelate ${\mathrm{La}}_{4}{\mathrm{Ni}}_{3}{\mathrm{O}}_{10}$ under pressure and its experimental examination: Comparison with ${\mathrm{La}}_{3}{\mathrm{Ni}}_{2}{\mathrm{O}}_{7}$}},
  author = {Sakakibara, Hirofumi and Ochi, Masayuki and Nagata, Hibiki and Ueki, Yuta and Sakurai, Hiroya and Matsumoto, Ryo and Terashima, Kensei and Hirose, Keisuke and Ohta, Hiroto and Kato, Masaki and Takano, Yoshihiko and Kuroki, Kazuhiko},
  journal = {Phys. Rev. B},
  volume = {109},
  issue = {14},
  pages = {144511},
  numpages = {10},
  year = {2024},
  month = {Apr},
  publisher = {American Physical Society},
  doi = {10.1103/PhysRevB.109.144511},
  url = {https://link.aps.org/doi/10.1103/PhysRevB.109.144511}
}

@article{Soni_2020,
  title = {{Flat bands and ferrimagnetic order in electronically correlated dice-lattice ribbons}},
  author = {Soni, Rahul and Kaushal, Nitin and Okamoto, Satoshi and Dagotto, Elbio},
  journal = {Phys. Rev. B},
  volume = {102},
  issue = {4},
  pages = {045105},
  numpages = {14},
  year = {2020},
  month = {Jul},
  publisher = {American Physical Society},
  doi = {10.1103/PhysRevB.102.045105},
  url = {https://link.aps.org/doi/10.1103/PhysRevB.102.045105}
}

@article{Sutherland_1986,
  title = {{Localization of electronic wave functions due to local topology}},
  author = {Sutherland, Bill},
  journal = {Phys. Rev. B},
  volume = {34},
  issue = {8},
  pages = {5208--5211},
  numpages = {0},
  year = {1986},
  month = {Oct},
  publisher = {American Physical Society},
  doi = {10.1103/PhysRevB.34.5208},
  url = {https://link.aps.org/doi/10.1103/PhysRevB.34.5208}
}

@article{Tanaka_2003,
  title = {{Stability of Ferromagnetism in the Hubbard Model on the Kagome Lattice}},
  author = {Tanaka, Akinori and Ueda, Hiromitsu},
  journal = {Phys. Rev. Lett.},
  volume = {90},
  issue = {6},
  pages = {067204},
  numpages = {4},
  year = {2003},
  month = {Feb},
  publisher = {American Physical Society},
  doi = {10.1103/PhysRevLett.90.067204},
  url = {https://link.aps.org/doi/10.1103/PhysRevLett.90.067204}
}

@article{Tasaki_1992,
  title = {{Ferromagnetism in the Hubbard models with degenerate single-electron ground states}},
  author = {Tasaki, Hal},
  journal = {Phys. Rev. Lett.},
  volume = {69},
  issue = {10},
  pages = {1608--1611},
  numpages = {0},
  year = {1992},
  month = {Sep},
  publisher = {American Physical Society},
  doi = {10.1103/PhysRevLett.69.1608},
  url = {https://link.aps.org/doi/10.1103/PhysRevLett.69.1608}
}

@article{Tasaki_1994,
  title = {{Stability of Ferromagnetism in the Hubbard Model}},
  author = {Tasaki, Hal},
  journal = {\emph{ibid.}},
  volume = {73},
  issue = {8},
  pages = {1158--1161},
  numpages = {0},
  year = {1994},
  month = {Aug},
  publisher = {American Physical Society},
  doi = {10.1103/PhysRevLett.73.1158},
  url = {https://link.aps.org/doi/10.1103/PhysRevLett.73.1158}
}

@article{Vidal_1998,
  title = {{Aharonov-Bohm Cages in Two-Dimensional Structures}},
  author = {Vidal, Julien and Mosseri, R\'emy and Dou\ifmmode \mbox{\c{c}}\else \c{c}\fi{}ot, Benoit},
  journal = {Phys. Rev. Lett.},
  volume = {81},
  issue = {26},
  pages = {5888--5891},
  numpages = {0},
  year = {1998},
  month = {Dec},
  publisher = {American Physical Society},
  doi = {10.1103/PhysRevLett.81.5888},
  url = {https://link.aps.org/doi/10.1103/PhysRevLett.81.5888}
}

@article{Wang_2011,
doi = {10.1209/0295-5075/93/57003},
url = {https://doi.org/10.1209/0295-5075/93/57003},
year = {2011},
month = {feb},
publisher = {},
volume = {93},
number = {5},
pages = {57003},
author = {Wang, Fa and Yang, Fan and Gao, Miao and Lu, Zhong-Yi and Xiang, Tao and Lee, Dung-Hai},
title = {{The electron pairing of K$_x$Fe$_{2-y}$Se$_2$}},
journal = {Europhys. Lett.},
}

@article{Yagi_2024,
  title = {{Theoretical analysis of the origin of the double-well band dispersion in the CuO double chains of ${\mathrm{Pr}}_{2}{\mathrm{Ba}}_{4}{\mathrm{Cu}}_{7}{\mathrm{O}}_{15\ensuremath{-}\ensuremath{\delta}}$ and its impact on superconductivity}},
  author = {Yagi, Toshiki and Ochi, Masayuki and Kuroki, Kazuhiko},
  journal = {Phys. Rev. B},
  volume = {110},
  issue = {18},
  pages = {184516},
  numpages = {9},
  year = {2024},
  month = {Nov},
  publisher = {American Physical Society},
  doi = {10.1103/PhysRevB.110.184516},
  url = {https://link.aps.org/doi/10.1103/PhysRevB.110.184516}
}

@Article{Zhou_2021,
author={Zhou, Haoxin and Xie, Tian and Taniguchi, Takashi and Watanabe, Kenji and Young, Andrea F.},
title={{Superconductivity in rhombohedral trilayer graphene}},
journal={Nature (London)},
year={2021},
month={Oct},
day={01},
volume={598},
number={7881},
pages={434-438},
issn={1476-4687},
doi={10.1038/s41586-021-03926-0},
url={https://doi.org/10.1038/s41586-021-03926-0}
}

@article{Raoux_2014,
  title = {{From Dia- to Paramagnetic Orbital Susceptibility of Massless Fermions}},
  author = {Raoux, A. and Morigi, M. and Fuchs, J.-N. and Pi\'echon, F. and Montambaux, G.},
  journal = {Phys. Rev. Lett.},
  volume = {112},
  issue = {2},
  pages = {026402},
  numpages = {5},
  year = {2014},
  month = {Jan},
  publisher = {American Physical Society},
  doi = {10.1103/PhysRevLett.112.026402},
  url = {https://link.aps.org/doi/10.1103/PhysRevLett.112.026402}
}

@Article{Abdi_2024,
author={Abdi, Mona
and Astinchap, Bandar},
title={{Exploration of a New $\alpha$-T3 System for Electronic Heat Capacity and Pauli Magnetic Susceptibility Properties via the Kane-Mele and Hubbard Model}},
journal={J. Phys. Chem. C},
year={2024},
month={Sep},
day={12},
publisher={American Chemical Society},
volume={128},
number={36},
pages={15179-15185},
issn={1932-7447},
doi={10.1021/acs.jpcc.4c04049},
url={https://doi.org/10.1021/acs.jpcc.4c04049}
}

@article{Alam_2019,
doi = {10.1088/1361-648X/ab3bf6},
url = {https://doi.org/10.1088/1361-648X/ab3bf6},
year = {2019},
month = {sep},
publisher = {IOP Publishing},
volume = {31},
number = {48},
pages = {485303},
author = {Alam, Mir Waqas and Souayeh, Basma and Islam, SK Firoz},
title = {{Enhancement of thermoelectric performance of a nanoribbon made of $\alpha$-$\mathcal{T}_3$ lattice}},
journal = {J. Phys.: Condens. Matter},
}

@article{Biswas_2016,
doi = {10.1088/0953-8984/28/49/495302},
url = {https://doi.org/10.1088/0953-8984/28/49/495302},
year = {2016},
month = {oct},
publisher = {IOP Publishing},
volume = {28},
number = {49},
pages = {495302},
author = {Biswas, Tutul and Kanti Ghosh, Tarun},
title = {{Magnetotransport properties of the $\alpha$-$T_3$ model}},
journal = {J. Phys.: Condens. Matter},
}

@article{YR-Chen_2019,
  title = {{Enhanced magneto-optical response due to the flat band in nanoribbons made from the $\ensuremath{\alpha}\ensuremath{-}{T}_{3}$ lattice}},
  author = {Chen, Yan-Ru and Xu, Yong and Wang, Jun and Liu, Jun-Feng and Ma, Zhongshui},
  journal = {Phys. Rev. B},
  volume = {99},
  issue = {4},
  pages = {045420},
  numpages = {7},
  year = {2019},
  month = {Jan},
  publisher = {American Physical Society},
  doi = {10.1103/PhysRevB.99.045420},
  url = {https://link.aps.org/doi/10.1103/PhysRevB.99.045420}
}

@article{Day_2019,
  title = {{Floquet topological phase transition in the $\ensuremath{\alpha}\text{\ensuremath{-}}{\mathcal{T}}_{3}$ lattice}},
  author = {Dey, Bashab and Ghosh, Tarun Kanti},
  journal = {Phys. Rev. B},
  volume = {99},
  issue = {20},
  pages = {205429},
  numpages = {9},
  year = {2019},
  month = {May},
  publisher = {American Physical Society},
  doi = {10.1103/PhysRevB.99.205429},
  url = {https://link.aps.org/doi/10.1103/PhysRevB.99.205429}
}

@article{Honerkamp_2008,
  title = {{Density Waves and Cooper Pairing on the Honeycomb Lattice}},
  author = {Honerkamp, Carsten},
  journal = {Phys. Rev. Lett.},
  volume = {100},
  issue = {14},
  pages = {146404},
  numpages = {4},
  year = {2008},
  month = {Apr},
  publisher = {American Physical Society},
  doi = {10.1103/PhysRevLett.100.146404},
  url = {https://link.aps.org/doi/10.1103/PhysRevLett.100.146404}
}

@article{Illes_2015,
  title = {{Hall quantization and optical conductivity evolution with variable Berry phase in the $\ensuremath{\alpha}\text{\ensuremath{-}}{T}_{3}$ model}},
  author = {Illes, E. and Carbotte, J. P. and Nicol, E. J.},
  journal = {Phys. Rev. B},
  volume = {92},
  issue = {24},
  pages = {245410},
  numpages = {9},
  year = {2015},
  month = {Dec},
  publisher = {American Physical Society},
  doi = {10.1103/PhysRevB.92.245410},
  url = {https://link.aps.org/doi/10.1103/PhysRevB.92.245410}
}

@article{Illes_2016,
  title = {{Magnetic properties of the $\ensuremath{\alpha}\ensuremath{-}{T}_{3}$ model: Magneto-optical conductivity and the Hofstadter butterfly}},
  author = {Illes, E. and Nicol, E. J.},
  journal = {Phys. Rev. B},
  volume = {94},
  issue = {12},
  pages = {125435},
  numpages = {13},
  year = {2016},
  month = {Sep},
  publisher = {American Physical Society},
  doi = {10.1103/PhysRevB.94.125435},
  url = {https://link.aps.org/doi/10.1103/PhysRevB.94.125435}
}

@article{Illes_2017,
  title = {{Klein tunneling in the $\ensuremath{\alpha}\text{\ensuremath{-}}{T}_{3}$ model}},
  author = {Illes, E. and Nicol, E. J.},
  journal = {Phys. Rev. B},
  volume = {95},
  issue = {23},
  pages = {235432},
  numpages = {8},
  year = {2017},
  month = {Jun},
  publisher = {American Physical Society},
  doi = {10.1103/PhysRevB.95.235432},
  url = {https://link.aps.org/doi/10.1103/PhysRevB.95.235432}
}

@article{Iurov_2019,
  title = {{Peculiar electronic states, symmetries, and Berry phases in irradiated $\ensuremath{\alpha}\text{\ensuremath{-}}{\mathrm{T}}_{3}$ materials}},
  author = {Iurov, Andrii and Gumbs, Godfrey and Huang, Danhong},
  journal = {Phys. Rev. B},
  volume = {99},
  issue = {20},
  pages = {205135},
  numpages = {20},
  year = {2019},
  month = {May},
  publisher = {American Physical Society},
  doi = {10.1103/PhysRevB.99.205135},
  url = {https://link.aps.org/doi/10.1103/PhysRevB.99.205135}
}

@article{Iurov_2020,
  title = {{Quantum-statistical theory for laser-tuned transport and optical conductivities of dressed electrons in $\ensuremath{\alpha}\ensuremath{-}{\mathcal{T}}_{3}$ materials}},
  author = {Iurov, Andrii and Zhemchuzhna, Liubov and Dahal, Dipendra and Gumbs, Godfrey and Huang, Danhong},
  journal = {Phys. Rev. B},
  volume = {101},
  issue = {3},
  pages = {035129},
  numpages = {12},
  year = {2020},
  month = {Jan},
  publisher = {American Physical Society},
  doi = {10.1103/PhysRevB.101.035129},
  url = {https://link.aps.org/doi/10.1103/PhysRevB.101.035129}
}

@article{Iurov_2023,
  title = {{Optical conductivity of gapped $\ensuremath{\alpha}\text{\ensuremath{-}}{\mathcal{T}}_{3}$ materials with a deformed flat band}},
  author = {Iurov, Andrii and Zhemchuzhna, Liubov and Gumbs, Godfrey and Huang, Danhong},
  journal = {Phys. Rev. B},
  volume = {107},
  issue = {19},
  pages = {195137},
  numpages = {14},
  year = {2023},
  month = {May},
  publisher = {American Physical Society},
  doi = {10.1103/PhysRevB.107.195137},
  url = {https://link.aps.org/doi/10.1103/PhysRevB.107.195137}
}

@article{KW-Lee_2024,
  title = {{Interplay between Haldane and modified Haldane models in $\ensuremath{\alpha}\text{\ensuremath{-}}{T}_{3}$ lattice: Band structures, phase diagrams, and edge states}},
  author = {Lee, Kok Wai and Fu, Pei-Hao and Ang, Yee Sin},
  journal = {Phys. Rev. B},
  volume = {109},
  issue = {23},
  pages = {235105},
  numpages = {11},
  year = {2024},
  month = {Jun},
  publisher = {American Physical Society},
  doi = {10.1103/PhysRevB.109.235105},
  url = {https://link.aps.org/doi/10.1103/PhysRevB.109.235105}
}

@article{KW-Lee_2025,
  title = {Floquet engineering of topological phase transitions in a quantum spin Hall $\ensuremath{\alpha}\text{\ensuremath{-}}{T}_{3}$ system},
  author = {Lee, Kok Wai and Calderon, Mateo Jalen Andrew and Yu, Xiang-Long and Lee, Ching Hua and Ang, Yee Sin and Fu, Pei-Hao},
  journal = {Phys. Rev. B},
  volume = {111},
  issue = {4},
  pages = {045406},
  numpages = {12},
  year = {2025},
  month = {Jan},
  publisher = {American Physical Society},
  doi = {10.1103/PhysRevB.111.045406},
  url = {https://link.aps.org/doi/10.1103/PhysRevB.111.045406}
}

@Article{F-Li_2022,
author={Li, Fu and Zhang, Qingtian and Chan, Kwok Sum},
title={{Novel transport properties of the $\alpha$-T$_3$ lattice with uniform electric and magnetic fields}},
journal={Sci. Rep.},
year={2022},
month={Jul},
day={29},
volume={12},
number={1},
pages={12987},
issn={2045-2322},
doi={10.1038/s41598-022-17288-8},
url={https://doi.org/10.1038/s41598-022-17288-8}
}

@article{R-Li_2023,
  title = {{Topological ac charge current and continuous invariant in the $\ensuremath{\alpha}\text{\ensuremath{-}}{T}_{3}$ lattice under a periodically varying strain}},
  author = {Li, Ruigang and Liu, Jun-Feng and Wang, Jun},
  journal = {Phys. Rev. B},
  volume = {108},
  issue = {11},
  pages = {115403},
  numpages = {8},
  year = {2023},
  month = {Sep},
  publisher = {American Physical Society},
  doi = {10.1103/PhysRevB.108.115403},
  url = {https://link.aps.org/doi/10.1103/PhysRevB.108.115403}
}

@article{SQ-Lin_2023,
title = {{Interaction-driven Chern insulating phases in the $\alpha$-$T_3$ lattice with Rashba spin-orbit coupling}},
journal = {iScience},
volume = {26},
number = {9},
pages = {107546},
year = {2023},
issn = {2589-0042},
doi = {https://doi.org/10.1016/j.isci.2023.107546},
url = {https://www.sciencedirect.com/science/article/pii/S2589004223016231},
author = {Shi-Qing Lin and Hui Tan and Pei-Hao Fu and Jun-Feng Liu},
}

@article{HL-Liu_2023,
  title = {{Chiral zero-energy modes in the disordered $\ensuremath{\alpha}\text{\ensuremath{-}}{T}_{3}$ lattice}},
  author = {Liu, Han-Lin and Wang, J. and Liu, Jun-Feng},
  journal = {Phys. Rev. B},
  volume = {107},
  issue = {12},
  pages = {125412},
  numpages = {8},
  year = {2023},
  month = {Mar},
  publisher = {American Physical Society},
  doi = {10.1103/PhysRevB.107.125412},
  url = {https://link.aps.org/doi/10.1103/PhysRevB.107.125412}
}

@Article{Mohanta_2023,
author={Mohanta, Narayan and Soni, Rahul and Okamoto, Satoshi and Dagotto, Elbio},
title={{Majorana corner states on the dice lattice}},
journal={Commun. Phys},
year={2023},
month={Sep},
day={06},
volume={6},
number={1},
pages={240},
issn={2399-3650},
doi={10.1038/s42005-023-01356-0},
url={https://doi.org/10.1038/s42005-023-01356-0}
}

@article{Parui_2024,
  title = {{Topological properties of nearly flat bands in bilayer $\ensuremath{\alpha}\ensuremath{-}{T}_{3}$ lattice}},
  author = {Parui, Puspita and Ghosh, Sovan and Chittari, Bheema Lingam},
  journal = {Phys. Rev. B},
  volume = {109},
  issue = {16},
  pages = {165118},
  numpages = {10},
  year = {2024},
  month = {Apr},
  publisher = {American Physical Society},
  doi = {10.1103/PhysRevB.109.165118},
  url = {https://link.aps.org/doi/10.1103/PhysRevB.109.165118}
}

@article{Piechon_2015,
doi = {10.1088/1742-6596/603/1/012001},
url = {https://doi.org/10.1088/1742-6596/603/1/012001},
year = {2015},
month = {mar},
publisher = {IOP Publishing},
volume = {603},
number = {1},
pages = {012001},
author = {Pi\'{e}chon, F and Fuchs, J-N and Raoux, A and Montambaux, G},
title = {{Tunable orbital susceptibility in $\alpha$-$T_3$ tight-binding models}},
journal = {J. Phys.: Conf. Ser.},
}

@article{Saleem_2025,
  title = {{Thermal transport properties of magnons on the $\ensuremath{\alpha}\text{\ensuremath{-}}{T}_{3}$ lattice}},
  author = {Saleem, Luqman and Abdullah, Hasan M. and Schwingenschl\"ogl, Udo and Manchon, Aur\'elien},
  journal = {Phys. Rev. B},
  volume = {111},
  issue = {2},
  pages = {024416},
  numpages = {10},
  year = {2025},
  month = {Jan},
  publisher = {American Physical Society},
  doi = {10.1103/PhysRevB.111.024416},
  url = {https://link.aps.org/doi/10.1103/PhysRevB.111.024416}
}

@article{Tamang_2023,
  title = {{Probing topological signatures in an optically driven $\ensuremath{\alpha}\ensuremath{-}{T}_{3}$ lattice}},
  author = {Tamang, Lakpa and Biswas, Tutul},
  journal = {Phys. Rev. B},
  volume = {107},
  issue = {8},
  pages = {085408},
  numpages = {12},
  year = {2023},
  month = {Feb},
  publisher = {American Physical Society},
  doi = {10.1103/PhysRevB.107.085408},
  url = {https://link.aps.org/doi/10.1103/PhysRevB.107.085408}
}

@article{H-Tan_2020,
doi = {10.1088/1361-6463/abcbbd},
url = {https://doi.org/10.1088/1361-6463/abcbbd},
year = {2020},
month = {dec},
publisher = {IOP Publishing},
volume = {54},
number = {10},
pages = {105303},
author = {Tan, Hui and Xu, Yong and Wang, Jun and Liu, Jun-Feng and Ma, Zhongshui},
title = {{Valley filter and giant magnetoresistance in zigzag $\alpha$-$T_3$ nanoribbons}},
journal = {J. Phys. D: Appl. Phys.},
}

@article{F-Wang_2011,
  title = {{Nearly flat band with Chern number $C=2$ on the dice lattice}},
  author = {Wang, Fa and Ran, Ying},
  journal = {Phys. Rev. B},
  volume = {84},
  issue = {24},
  pages = {241103},
  numpages = {5},
  year = {2011},
  month = {Dec},
  publisher = {American Physical Society},
  doi = {10.1103/PhysRevB.84.241103},
  url = {https://link.aps.org/doi/10.1103/PhysRevB.84.241103}
}

@article{JJ-Wang_2020,
  title = {Integer quantum Hall effect of the $\ensuremath{\alpha}\text{\ensuremath{-}}{T}_{3}$ model with a broken flat band},
  author = {Wang, Juan Juan and Liu, S. and Wang, J. and Liu, Jun-Feng},
  journal = {Phys. Rev. B},
  volume = {102},
  issue = {23},
  pages = {235414},
  numpages = {10},
  year = {2020},
  month = {Dec},
  publisher = {American Physical Society},
  doi = {10.1103/PhysRevB.102.235414},
  url = {https://link.aps.org/doi/10.1103/PhysRevB.102.235414}
}

@article{J-Wang_2021,
  title = {{Quantum spin Hall phase transition in the $\ensuremath{\alpha}\text{\ensuremath{-}}{T}_{3}$ lattice}},
  author = {Wang, J. and Liu, Jun-Feng},
  journal = {Phys. Rev. B},
  volume = {103},
  issue = {7},
  pages = {075419},
  numpages = {7},
  year = {2021},
  month = {Feb},
  publisher = {American Physical Society},
  doi = {10.1103/PhysRevB.103.075419},
  url = {https://link.aps.org/doi/10.1103/PhysRevB.103.075419}
}

@article{YJ-Wei_2024,
doi = {10.1088/1674-1056/ad6f91},
url = {https://doi.org/10.1088/1674-1056/ad6f91},
year = {2024},
month = {nov},
publisher = {Chinese Physical Society and IOP Publishing Ltd},
volume = {33},
number = {11},
pages = {117201},
author = {Wei, Ya-Jun and Wang, Jun},
title = {{Valley switch effect in an $\alpha$-$T_3$ lattice-based superconducting interferometer}},
journal = {Chin. Phys. B},
}

@article{X-Zhou_2021,
  title = {{Andreev reflection and Josephson effect in the $\ensuremath{\alpha}\ensuremath{-}{T}_{3}$ lattice}},
  author = {Zhou, Xingfei},
  journal = {Phys. Rev. B},
  volume = {104},
  issue = {12},
  pages = {125441},
  numpages = {9},
  year = {2021},
  month = {Sep},
  publisher = {American Physical Society},
  doi = {10.1103/PhysRevB.104.125441},
  url = {https://link.aps.org/doi/10.1103/PhysRevB.104.125441}
}

@Article{Orlita_2014,
author={Orlita, M. and Basko, D. M. and Zholudev, M. S. and Teppe, F. and Knap, W. and Gavrilenko, V. I. and Mikhailov, N. N. and Dvoretskii, S. A. and Neugebauer, P. and Faugeras, C. and Barra, A.-L. and Martinez, G. and Potemski, M.},
title={{Observation of three-dimensional massless Kane fermions in a zinc-blende crystal}},
journal={Nat. Phys.},
year={2014},
month={Mar},
day={01},
volume={10},
number={3},
pages={233-238},
issn={1745-2481},
doi={10.1038/nphys2857},
url={https://doi.org/10.1038/nphys2857}
}

@article{Malcolm_2015,
  title = {{Magneto-optics of massless Kane fermions: Role of the flat band and unusual Berry phase}},
  author = {Malcolm, J. D. and Nicol, E. J.},
  journal = {Phys. Rev. B},
  volume = {92},
  issue = {3},
  pages = {035118},
  numpages = {4},
  year = {2015},
  month = {Jul},
  publisher = {American Physical Society},
  doi = {10.1103/PhysRevB.92.035118},
  url = {https://link.aps.org/doi/10.1103/PhysRevB.92.035118}
}

@article{Rizzi_2006,
  title = {{Phase diagram of the Bose-Hubbard model with ${\mathcal{T}}_{3}$ symmetry}},
  author = {Rizzi, Matteo and Cataudella, Vittorio and Fazio, Rosario},
  journal = {Phys. Rev. B},
  volume = {73},
  issue = {14},
  pages = {144511},
  numpages = {15},
  year = {2006},
  month = {Apr},
  publisher = {American Physical Society},
  doi = {10.1103/PhysRevB.73.144511},
  url = {https://link.aps.org/doi/10.1103/PhysRevB.73.144511}
}

@article{Bercioux_2009,
  title = {{Massless Dirac-Weyl fermions in a ${\mathcal{T}}_{3}$ optical lattice}},
  author = {Bercioux, D. and Urban, D. F. and Grabert, H. and H\"ausler, W.},
  journal = {Phys. Rev. A},
  volume = {80},
  issue = {6},
  pages = {063603},
  numpages = {4},
  year = {2009},
  month = {Dec},
  publisher = {American Physical Society},
  doi = {10.1103/PhysRevA.80.063603},
  url = {https://link.aps.org/doi/10.1103/PhysRevA.80.063603}
}

@article{Okamoto_2025_review,
  title = {{Novel phenomena in transition-metal oxide thin films and heterostructures with strong correlations and spin-orbit coupling}},
  author = {Okamoto, Satoshi and Mohanta, Narayan and Lee, Ho Nyung and Moreo, Adriana and Dagotto, Elbio},
  journal = {Phys. Rev. Mater.},
  volume = {9},
  issue = {5},
  pages = {050301},
  numpages = {18},
  year = {2025},
  month = {May},
  publisher = {American Physical Society},
  doi = {10.1103/PhysRevMaterials.9.050301},
  url = {https://link.aps.org/doi/10.1103/PhysRevMaterials.9.050301}
}

@Article{Geng_2026,
author={Geng, Songyuan and Wang, Xin and Guo, Risi and Qiu, Chen and Chen, Fangjie and Wang, Qun and Li, Kangjie and Hao, Peipei and Liang, Hanpu and Huang, Yang and Wu, Yunbo and Cui, Shengtao and Sun, Zhe and Kim, Timur K. and Cacho, Cephise and Dessau, Daniel S. and Zhou, Benjamin T. and Li, Haoxiang},
title={{Experimental realization of dice-lattice flat band at the Fermi level in layered electride YCl}},
journal={Nat. Commun.},
year={2026},
month={Jan},
day={31},
volume={17},
number={1},
pages={2213},
issn={2041-1723},
doi={10.1038/s41467-026-69049-0},
url={https://doi.org/10.1038/s41467-026-69049-0}
}

@misc{Mojarro_arxiv,
      title={{Superconductivity and geometric superfluid weight of a tunable flat band system}}, 
      author={M. A. Mojarro and Sergio E. Ulloa},
      eprint={2512.09901},
      archivePrefix={arXiv}, 
}

@article{Gorbar_2021,
  title = {{Gap generation and flat band catalysis in dice model with local interaction}},
  author = {Gorbar, E. V. and Gusynin, V. P. and Oriekhov, D. O.},
  journal = {Phys. Rev. B},
  volume = {103},
  issue = {15},
  pages = {155155},
  numpages = {12},
  year = {2021},
  month = {Apr},
  publisher = {American Physical Society},
  doi = {10.1103/PhysRevB.103.155155},
  url = {https://link.aps.org/doi/10.1103/PhysRevB.103.155155}
}

@article{YP-Lin_2018,
  title = {{Exotic superconductivity with enhanced energy scales in materials with three band crossings}},
  author = {Lin, Yu-Ping and Nandkishore, Rahul M.},
  journal = {Phys. Rev. B},
  volume = {97},
  issue = {13},
  pages = {134521},
  numpages = {15},
  year = {2018},
  month = {Apr},
  publisher = {American Physical Society},
  doi = {10.1103/PhysRevB.97.134521},
  url = {https://link.aps.org/doi/10.1103/PhysRevB.97.134521}
}

@article{YP-Lin_2020,
  title = {{Chiral flat band superconductivity from symmetry-protected three-band crossings}},
  author = {Lin, Yu-Ping},
  journal = {Phys. Rev. Res.},
  volume = {2},
  issue = {4},
  pages = {043209},
  numpages = {9},
  year = {2020},
  month = {Nov},
  publisher = {American Physical Society},
  doi = {10.1103/PhysRevResearch.2.043209},
  url = {https://link.aps.org/doi/10.1103/PhysRevResearch.2.043209}
}

@article{Mandal_2021a,
  title = {{$p$-wave superconductivity and the axiplanar phase of triple-point fermions}},
  author = {Mandal, Subrata and Herbut, Igor F.},
  journal = {Phys. Rev. B},
  volume = {104},
  issue = {18},
  pages = {L180507},
  numpages = {5},
  year = {2021},
  month = {Nov},
  publisher = {American Physical Society},
  doi = {10.1103/PhysRevB.104.L180507},
  url = {https://link.aps.org/doi/10.1103/PhysRevB.104.L180507}
}

@article{Mandal_2021b,
  title = {{Time-reversal symmetry breaking and $d$-wave superconductivity of triple-point fermions}},
  author = {Mandal, Subrata and Link, Julia M. and Herbut, Igor F.},
  journal = {Phys. Rev. B},
  volume = {104},
  issue = {13},
  pages = {134512},
  numpages = {12},
  year = {2021},
  month = {Oct},
  publisher = {American Physical Society},
  doi = {10.1103/PhysRevB.104.134512},
  url = {https://link.aps.org/doi/10.1103/PhysRevB.104.134512}
}

@Article{Sim_2022,
author={Sim, GiBaik and Park, Moon Jip and Lee, SungBin},
title={Topological triplet-superconductivity in spin-1 semimetal},
journal={Commun. Phys.},
year={2022},
month={Sep},
day={07},
volume={5},
number={1},
pages={220},
issn={2399-3650},
doi={10.1038/s42005-022-00992-2},
url={https://doi.org/10.1038/s42005-022-00992-2}
}

@article{Zyuzin_2022,
  title = {{Preformed Cooper pairs in flat-band semimetals}},
  author = {Zyuzin, Alexander A. and Zyuzin, A. Yu.},
  journal = {Phys. Rev. B},
  volume = {106},
  issue = {2},
  pages = {L020502},
  numpages = {5},
  year = {2022},
  month = {Jul},
  publisher = {American Physical Society},
  doi = {10.1103/PhysRevB.106.L020502},
  url = {https://link.aps.org/doi/10.1103/PhysRevB.106.L020502}
}

@article{Yoshida_2025,
  title = {{Chirality detection through vortex bound states in a ($d+i{d}^{\ensuremath{'}}$)-wave superconductor}},
  author = {Yoshida, Soma and Tanaka, Yukio and Golubov, Alexander A. and Suzuki, Shu-Ichiro},
  journal = {Phys. Rev. B},
  volume = {111},
  issue = {1},
  pages = {014511},
  numpages = {10},
  year = {2025},
  month = {Jan},
  publisher = {American Physical Society},
  doi = {10.1103/PhysRevB.111.014511},
  url = {https://link.aps.org/doi/10.1103/PhysRevB.111.014511}
}

@article{Huang_2014,
  title = {{Vanishing edge currents in non-$p$-wave topological chiral superconductors}},
  author = {Huang, Wen and Taylor, Edward and Kallin, Catherine},
  journal = {Phys. Rev. B},
  volume = {90},
  issue = {22},
  pages = {224519},
  numpages = {9},
  year = {2014},
  month = {Dec},
  publisher = {American Physical Society},
  doi = {10.1103/PhysRevB.90.224519},
  url = {https://link.aps.org/doi/10.1103/PhysRevB.90.224519}
}

@article{Cadorim_2024,
  title = {{Vortical versus skyrmionic states in the topological phase of a twisted bilayer with $d$-wave superconducting pairing}},
  author = {Cadorim, Leonardo R. and Sardella, Edson and Milo\ifmmode \check{s}\else \v{s}\fi{}evi\ifmmode \acute{c}\else \'{c}\fi{}, Milorad V.},
  journal = {Phys. Rev. B},
  volume = {110},
  issue = {6},
  pages = {064508},
  numpages = {11},
  year = {2024},
  month = {Aug},
  publisher = {American Physical Society},
  doi = {10.1103/PhysRevB.110.064508},
  url = {https://link.aps.org/doi/10.1103/PhysRevB.110.064508}
}

\end{document}